\documentclass[titlepage,11pt]{article}

\usepackage[dvips]{graphicx}
\usepackage[myheadings]{fullpage}
\usepackage{pmetrika}
\usepackage{pmbib}
\usepackage[natbibapa]{apacite}
\usepackage{submit}
\usepackage{graphics,graphicx}
\usepackage{longtable}
\usepackage{wrapfig}
\usepackage{amsmath,verbatim,amssymb,epsfig,color,lscape}
\usepackage{fancyhdr}
\usepackage{lastpage}
\usepackage{threeparttable}
\usepackage{multirow}
\newcommand{\hide}[1]{}
\usepackage[normalem]{ulem}
\usepackage{bm}
\newcommand{\s}{\vspace{.25cm}}

\usepackage{tikz}
\usetikzlibrary{bayesnet}
\usetikzlibrary{fit,positioning}

\newcommand{\YY}{\bm{Y}}
\newcommand{\yy}{\mbox{$\mathbf y$}}

\newcommand{\be}{\mbox{\boldmath $\beta$}}

\def\th{\mbox{\boldmath $\alpha$}}

\def\logit{\mbox{\rm logit}}

\newcommand{\bi}{\begin{itemize}}
\newcommand{\ei}{\end{itemize}}

\newcommand{\mR}{\mathbb{R}}
\newcommand{\msim}{\mathop{\rm \sim}}
\newcommand{\ind}{\msim\limits^{\mbox{\tiny ind}}}
\newcommand{\iid}{\msim\limits^{\mbox{\tiny iid}}}

\begin{document}

\begin{titlepage}

\title{Mapping unobserved item-respondent interactions:\newline
A latent space item response model with interaction map}

\author{Minjeong Jeon}
\affil{University of California, Los Angeles}

\author{Ick Hoon Jin}
\affil{Yonsei University}

\author{Michael Schweinberger}
\affil{Rice University}

\author{Samuel Baugh}
\affil{University of California, Los Angeles}

\vspace{\fill}\centerline{\today}\vspace{\fill}

\contact{Ick Hoon Jin was partially supported by the Yonsei University Research Fund 2019-22-0210 and the Basic Science Research Program through the National Research Foundation of Korea (NRF 2020R1A2C1A01009881). 
Michael Schweinberger was partially supported by NSF award DMS-1812119.

Correspondence should be addressed to Minjeong Jeon, 
Graduate School of Education and Information Studies,
University of California, Los Angeles, CA 90095, USA. 
E-Mail: mjjeon@ucla.edu. 
Jeon, Jin, and Schweinberger are co-first authors with equal contribution. }

\end{titlepage}

\setcounter{page}{2}

\vspace*{2\baselineskip}

\RepeatTitle{Mapping unobserved item-respondent interactions: A latent space item response model with interaction map}\vskip3pt

\linespacing{1.5}

\abstracthead

\begin{abstract}
Classic item response models assume that all items with the same difficulty have the same response probability among  all respondents with the same ability. These assumptions, however, may very well be violated in practice, and it is not straightforward to assess whether these assumptions are violated, because neither the abilities of respondents nor the difficulties of items are observed. An example is an educational assessment where unobserved heterogeneity is present, arising from unobserved variables such as cultural background and upbringing of students, the quality of mentorship and other forms of emotional and professional support received by students, and other unobserved variables that may affect response probabilities. To address such violations of assumptions, we introduce a novel latent space model which assumes that both items and respondents are embedded in an unobserved metric space, with the probability of a correct response decreasing as a function of the distance between the respondent's and the item's position in the latent space. The resulting latent space approach provides an interaction map that represents interactions of respondents and items, and helps derive insightful diagnostic information on items as well as respondents. In practice, such interaction maps enable teachers to detect students from underrepresented groups who need more support than other students. We provide empirical evidence to demonstrate the usefulness of the proposed latent space approach, along with simulation results.

\end{abstract}\vspace{\fill}

\pagebreak

\section{Introduction}\label{sec:intro}

Item response theory (IRT) is a widely used approach for analyzing responses to test items given by test takers,
called respondents. 
A classic IRT model,
the Rasch model \citep{Rasch:1961}, 
assumes that the log odds of the probability of a correct response $Y_{j,i} = 1$ to binary item $i$ by respondent $j$ is of the form
\begin{equation} \label{eq:1}
\mbox{logit} (\mathbb{P}(Y_{j,i} = 1 \mid \alpha_j,\; \beta_i)) \;=\; \alpha_j + \beta_i.     
\end{equation}
In words,
the probability of a correct response to item $i$ by respondent $j$ is a function of two attributes: 
one associated with respondent $j$,
$\alpha_j \in \mR$, 
and the other associated with item $i$,
$\beta_i \in \mR$. 
The main effect $\alpha_j$ represents the ability of respondent $j$,
while the main effect $\beta_i$ of item $i$ reveals how easily item $i$ is correctly answered. 

The Rasch model rests on the following assumptions:
(1) the item responses of any respondent are independent of the item responses of any other respondent,
conditional on the abilities of the respondents and the easiness levels of the items;
% (unit independence);
(2) for each respondent, 
the responses to items are independent,
conditional on the ability of the respondent and the easiness levels of the items;
% (local independence);
(3) for each item, 
respondents with the same level of ability have the same success probability;
% (unit homogeneity),
and, 
for each respondent,
items with the same easiness level have the same success probability.
% (item homogeneity). 

These assumptions, 
however, 
may very well be violated in practice:
e.g.,
in some educational assessments it is not credible that all items with the same easiness level have the same response probability for all respondents with the same ability.
An example is an educational assessment where unobserved heterogeneity is present,
arising from unobserved variables such as cultural background and upbringing of students, the quality of mentorship and other forms of emotional and professional support received by students, and other unobserved variables that may affect response probabilities.
Worse,
in practice it might be hard if not impossible to assess whether such assumptions are violated,
because the abilities of respondents and the easiness levels of items are unobserved.

\textcolor{black}{ 
To address violations of these assumptions,
we introduce a novel latent space model which assumes that both items and respondents are embedded in an unobserved metric space,
with the probability of a correct response decreasing as a function of the distance between the respondent's and the item's position in the latent space.
The resulting latent space approach provides an interaction map that represents interactions of respondents and items,
and helps derive insightful diagnostic information on items as well as respondents.
}

The novel latent space model we introduced is inspired by recent work on item response models that view item response data as networks.
For example, 
\citet{Bo08} described a network analysis of psychological constructs, 
where covariance between observed indicator variables stems from interactions among items.
% without using latent variables. 
More recently, 
\citet{Epskamp:2018} proposed a network approach based on Gaussian graphical models,
which can include latent variables. 
\citet{Marsman:2018} studied relations between an Ising model and other item response models. 
All of them are concerned with interactions among items,
not respondents.
Another recent development in network modeling of item response data is the doubly latent space joint model (DLSJM) of \citet{Jin:18} and its extension to hierarchical data \citep{Jin:2018b},
henceforth called the network item response model (NIRM).
\textcolor{black}{The NIRM approach is inspired by latent space models of network data \citep{Hoff:2002, Schweinberger:03, sewell2015latent, SmAsCa19}.
Latent space models of network data and non-network data may be viewed as a model-based alternative to multidimensional scaling (MDS),
having the advantage of enabling model-based statistical inference and capturing the uncertainty about the positions of units in the latent space,
in contrast to MDS.
The NIRM approach constructs functions of item response data which can be viewed as network data:
respondent-respondent networks consisting of links between respondents who both gave the correct response to an item (one network for each item),
and item-item networks consisting of links between items that received the correct response by a respondent (one network for each respondent).}
Our proposed approach is inspired by NIRM, 
but simpler than the NIRM approach.
We view item response data as a bipartite network, 
consisting of links between respondents on the one hand and items on the other hand. 
This change in perspective comes with important benefits,
including, 
but not limited to:
(1) We work with the original item response data rather than functions of item response data;
(2) we have a single network rather than multiple networks,
which would have to be combined;
(3) we can examine relationships between items and respondents without choosing a procedure that combines multiple networks,
which -- when the procedure is inappropriate -- introduces an additional source of error;
and (4) our approach is closely related to the Rasch model,
which facilitates interpretation.

Our paper is organized as follows.
We introduce latent space models in Section 2,
discuss Bayesian inference in Section 3,
and present examples along with simulation results in Sections 4 and 5. 
We conclude our paper in Section 6. 

\section{Model}\label{sec:model}

{\color{black}

\subsection{Latent Space Item Response Model} 
\label{development}

We consider item response data consisting of a binary $N$ by $I$ matrix $\YY \in \{0, 1\}^{N \times I}$, 
where $Y_{j,i} = 1$ indicates a correct response by respondent $j$ to item $i$,
whereas $Y_{j,i} = 0$ indicates an incorrect response.
Extensions to non-binary item response data are straightforward,
by replacing the logit-link function for binary item response data by a suitable link function for non-binary item response data,
as in generalized linear models \citep{MpNj83}.

To capture unobserved interactions of respondents and items,
we assume that both respondents and items are embedded in an unobserved metric space.
A metric space $(\mathbb{M}, d)$ consists of a space $\mathbb{M}$ and a distance function $d: \mathbb{M} \times \mathbb{M} \mapsto [0, +\infty)$ assigning distances to pairs of points $(\bm{a}, \bm{b}) \in \mathbb{M} \times \mathbb{M}$ (corresponding to positions of respondents and items),
which satisfy
\begin{itemize}
\item reflexivity: $d(\bm{a},\, \bm{b}) = 0$ if and only if $\bm{a} = \bm{b} \in \mathbb{M}$;
\item symmetry: $d(\bm{a},\, \bm{b}) = d(\bm{b}, \bm{a})$ for all $\bm{a},\, \bm{b} \in \mathbb{M}$;
\item triangle inequality: $d(\bm{a},\, \bm{b}) \leq  d(\bm{a}, \bm{c}) + d(\bm{b},\bm{c})$ for all $\bm{a},\, \bm{b},\, \bm{c} \in \mathbb{M}$.
\end{itemize}
We follow the convention in statistical network analysis \citep{Hoff:2002} and assume that $\mathbb{M}$ is $p$-dimensional Euclidean space $\mR^p$ with known dimension $p \geq 1$.
% although the main ideas of the proposed latent space approach are not restricted to Euclidean spaces.
Some possible choices of the distance function $d: \mR^p \times \mR^p \mapsto [0, +\infty)$ are:
\begin{itemize}
\item $\ell_1$-distance (city-block distance): $d(\bm{a},\, \bm{b}) =  ||\bm{a} - \bm{b}||_1 = \sum_{i=1}^p |a_i - b_i|$.
\item $\ell_2$-distance (Euclidean distance): $d(\bm{a},\, \bm{b}) = ||\bm{a} - \bm{b}||_2 = \sqrt{\sum_{i=1}^p (a_i - b_i)^2}$.
\item $\ell_\infty$-distance (maximum distance): $d(\bm{a},\, \bm{b}) =  ||\bm{a} - \bm{b}||_\infty = \max_{1 \leq i \leq p}\, |a_i - b_i|$.
\end{itemize}
To capture unobserved interactions of respondents and items,
we assume that the probability of a correct response by respondent $j$ to item $i$ depends on the position $\bm{a}_j \in \mR^p$ of respondent $j$ and the position $\bm{b}_i \in \mR^p$ of item $i$ in the shared metric space:
\begin{equation}\label{eq:2}
\mbox{logit} (\mathbb{P} (Y_{j,i} = 1 \mid \alpha_j,\, \beta_i,\, \bm{a}_j,\, \bm{b}_i)) \;=\; 
\alpha_j + \beta_i + g(\bm{a}_j,\; \bm{b}_i),
\end{equation}
where $g: \mR^p \times \mR^p \mapsto \mR$ is a real-valued function of the positions of respondent $j$ and item $i$.
There are many possible choices of the function $g$.
We discuss two natural choices:
\bi
\item multiplicative effect: $g(\bm{a}_j,\, \bm{b}_i) = \bm{a}_j^\top\, \bm{b}_i$,
where $\bm{a}_j^\top\, \bm{b}_i$ is the inner product of $\bm{a}_j$ and $\bm{b}_i$;
\item distance effect: $g(\bm{a}_j,\, \bm{b}_i) = - \gamma\; d(\bm{a}_j, \bm{b}_i)$,
where $d(\bm{a}_j,\, \bm{b}_i)$ is the distance between $\bm{a}_j$ and $\bm{b}_i$ (e.g., the $\ell_1$-distance, $\ell_2$-distance, or $\ell_\infty$-distance) and $\gamma \geq 0$ is the weight of the distance term;
note that $\gamma > 0$ ensures that increasing the distance decreases the probability of a correct response.
\ei
While both choices are legitimate and have advantages and disadvantages,
we believe that the distance effect is easier to interpret than the multiplicative effect.
For example,
the effect of the inner product on the log odds of a correct response is 0 when $\bm{a}_j$ and $\bm{b}_i$ are orthogonal,
regardless of whether the distance between $\bm{a}_j$ and $\bm{b}_i$ is small or large:
e.g.,
if $d$ is the $\ell_2$-distance and $\bm{a}_j = (0,\; 1/100)$ and $\bm{b}_i = (1/100,\, 0)$,
then $d(\bm{a}_j,\, \bm{b}_i) = 0.01$,
whereas $\bm{a}_j = (0,\; 100)$ and $\bm{b}_i = (100,\, 0)$ implies $d(\bm{a}_j,\, \bm{b}_i) = 141.42$.
In both examples,
$\bm{a}_j^\top\, \bm{b}_i = 0$,
but in the first case the distance between the two vectors is small whereas in the second case it is large.
Therefore,
to interpret interaction maps and the effect of interactions on the probability of a correct response under the multiplicative effects model,
one needs to pay careful attention to the angle of the vectors $\bm{a}_j$ and $\bm{b}_i$,
in addition to the lengths of $\bm{a}_j$ and $\bm{b}_i$.
That makes the resulting interaction maps more challenging to use by practitioners and applied researchers, %(e.g., teachers in the classroom),
undermining one of the main advantages of the latent space approach.
We therefore focus on the model with the distance effect,
although the multiplicative effects model would be an interesting alternative.

It is worth noting that the latent space model with $\gamma = 0$ is equivalent to the Rasch model,
so the latent space model with $\gamma \geq 0$ can be viewed as a generalization of the Rasch model.
In practice,
we determine whether $\gamma = 0$ or $\gamma > 0$ via model selection, as described in Section \ref{sec:model_selection}.  %by using spike-and-slab priors.
If $\gamma > 0$,
the latent space model has added value compared with the Rasch model.
The added value of the latent space model is that it captures deviations from the main effects $\alpha_j$ and $\beta_i$ of the Rasch model -- that is,
interactions of respondent $j$ and item $i$ -- and visualizes those interactions by embedding respondents along with items in a shared metric space.
As a consequence,
it is natural to interpret the metric space as an interaction map,
rather than an ability space.
 
We discuss below properties of the latent space model,
including practical and theoretical advantages along with a network view of item response data. 
Statistical issues -- including identifiability issues -- are discussed in Section \ref{sec:estimation}.
}

{\color{black}
\subsection{Properties}

\subsubsection{Latent Space Model as Network Model} 

The latent space model introduced above was inspired by latent space models of network data -- as mentioned in Section 1 -- and it may be viewed as a network model.
Specifically,
one may view item response data as a bipartite network \citep{WsFk94},
consisting of links between respondents on the one hand and items on the other hand,
where links correspond to correct responses by respondents to items.
In contrast to conventional network data,
bipartite respondent-item networks consist of two sets of units rather than one set of units (i.e., the set of respondents and the set of items).
The proposed modeling framework then assumes that the bipartite network was generated by a latent space model,
with both respondents and items embedded in a shared metric space.
% similar to latent space models of bipartite network data \citep[e.g.,][]{Friel:2016}.

It is worth noting that viewing the proposed latent space model as a network model may or may not be useful, 
although network models have turned out to be useful across a staggering number of fields -- not the least in artificial intelligence (AI), 
where deep neural networks have enabled substantial advances in voice recognition and computer vision \citep{Goodfellow-et-al-2016}.
In probability and statistics,
there are two main lines of research involving networks.
First,
graphical models use graphs to represent conditional independence structure \citep{Pe88,La96},
that is,
model structure.
Graphical models assume that the variables of interest (here: items) are the vertices of a graph,
and the absence of a link between two vertices indicates that the two corresponding variables are independent conditional on all other variables.
Links therefore indicate conditional dependencies,
given all other vertices.
Examples are Gaussian graphical models,
Ising models,
and Boltzmann machines in AI.
In the IRT literature, \citet{Epskamp:2018} and \citet{Marsman:2018} and others followed a graphical model approach to studying interactions among items (albeit not respondent-item interactions, as we do).
Second,
random graph models use graphs to represent data structure.
For example,
in social network analysis \citep{WsFk94},
individuals are the vertices of a graph,
and the links may indicate friendships among individuals.

The proposed latent space model can be represented as a graphical model (with variables $Y_{j,i}$, $\bm{a}_j$, and $\bm{b}_i$ constituting the vertices of a graph and links corresponding to conditional dependencies among these variables) or as a random graph model (with respondents $j$ and items $i$ constituting the vertices of a graph and links corresponding to correct responses).
Whether it is useful to view the proposed model as a graphical model or as a random graph model is open to discussion,
but -- regardless of whether one embraces a network view -- the proposed modeling framework has practical and theoretical advantages.

\subsubsection{Practical advantages}

A unique advantage of the proposed latent space approach is that it provides a geometric representation of interactions among respondents and items in a low-dimensional space, 
e.g.,
$\mR^2$.
The interaction structure mapped into two-dimensional Euclidean space helps detect unobserved characteristics of items and respondents.

\begin{figure}[t]
\centering
\begin{tabular}{cc}
(a) Rasch model &   (b) Rasch model with local dependence \\
\includegraphics[width=0.4\textwidth]{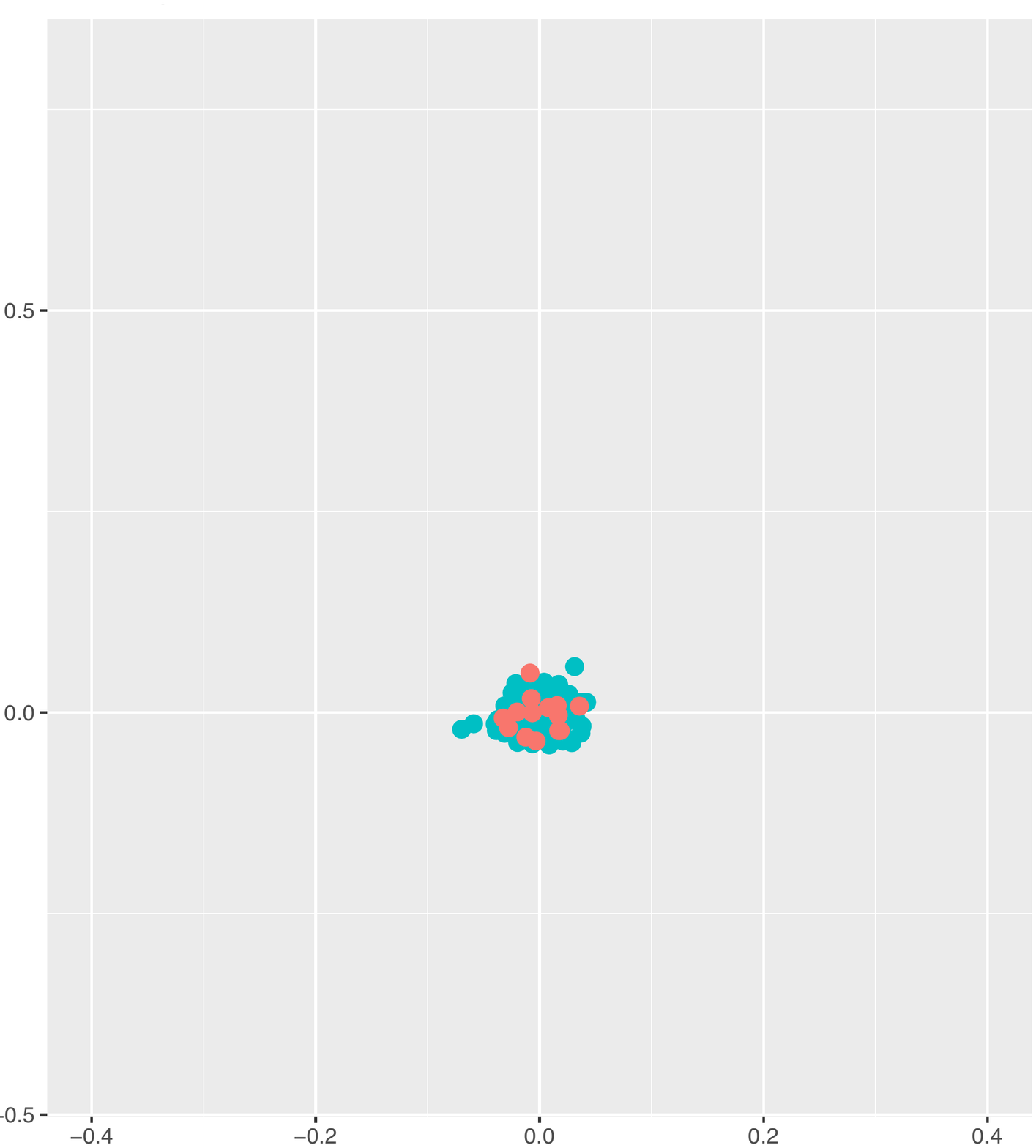} &
\includegraphics[width=0.4\textwidth]{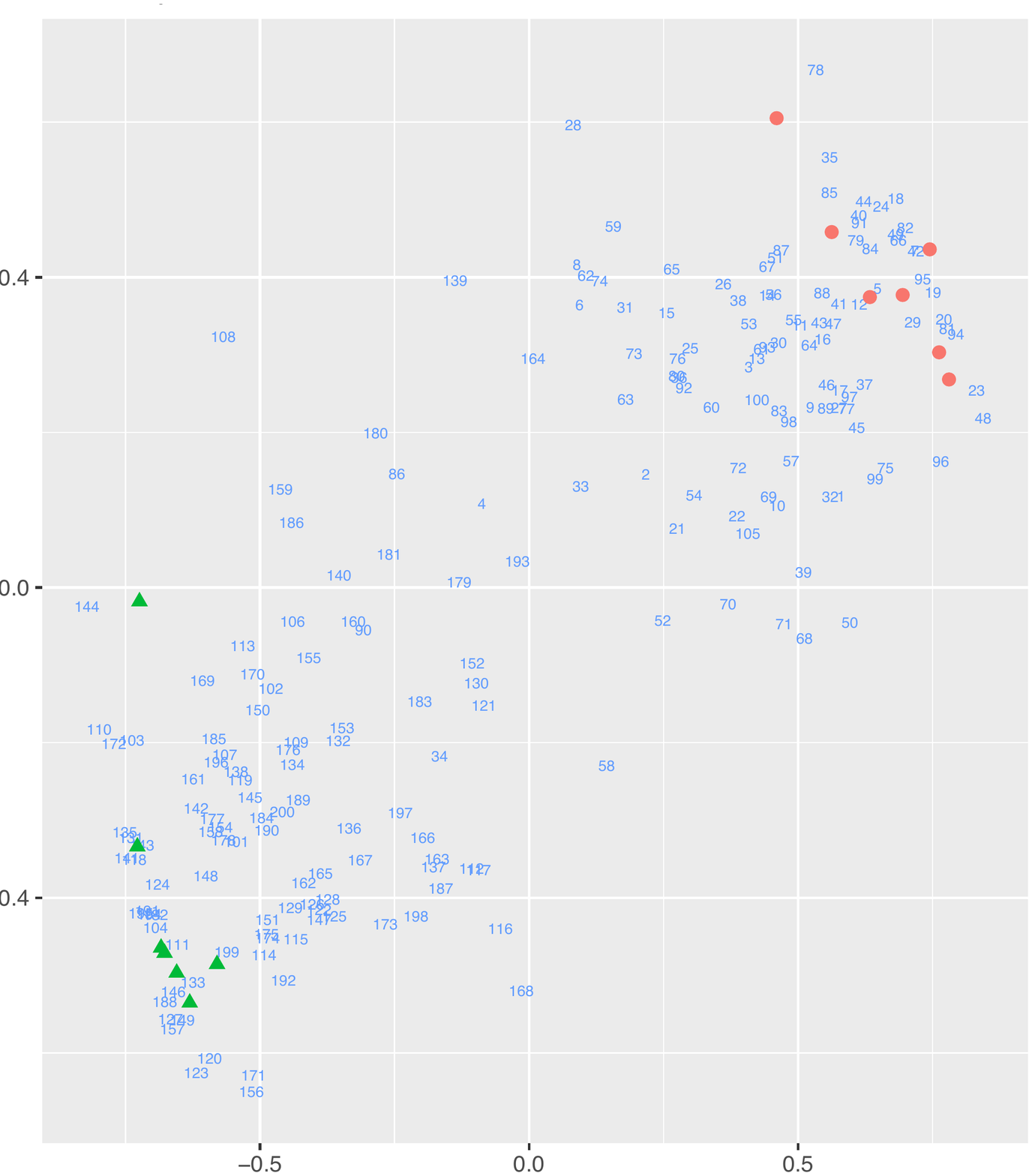} \\
\end{tabular}
\caption{\label{zw_random}
Estimated latent space configurations based on responses to 14 items by 200 respondents, 
generated by (a) the Rasch model and (b) the Rasch model with local dependence. 
In (a),
blue circles represent respondents and red circles represent items,
whereas in (b) blue numbers represent respondents,
red circles represent the first 7 items,
and green triangles represent the last 7 items.
%The first 100 people (1--100; group 1) have strong dependence with Items 1--7 (b\_I1), while the last 100 people (101-200; group 1) have strong dependence with Items 8--14 (b\_I2). 
%As expected, respondents 1 -- 100 are closer to the first 7 items than respondents 101 -- 200.
%,
%whereas respondents 101 -- 200 are close to the last 7 items.
}
\end{figure}

To demonstrate,
we conduct a simulation study.
The simulation results are based on binary responses to 14 items by 200 respondents.
First, 
data are generated from the Rasch model.
Second, 
data are generated from the Rasch model with local dependence,
that is,
the responses of the first 100 respondents to the first 7 items exhibit strong local dependence in the sense of \citet{chen:97},
and the responses of the last 100 respondents to the last 7 items likewise exhibit strong local dependence.  
The proposed latent space model with $p=2$ is estimated from both datasets, 
using the Bayesian Markov chain Monte Carlo algorithm described in Section \ref{sec:estimation}. 
Additional details are provided in Appendix A of the supplement. 
Figure \ref{zw_random} shows that in the first case all items and respondents are located close to the origin of $\mR^2$,
whereas in the second case the two groups of items are well-separated in $\mR^2$ and the two groups of 100 respondents are located close to the respective sets of items, 
as expected.

\paragraph{Latent space dimension.} 
Throughout the remainder of the paper,
we choose $\mathbb{M} = \mR^2$,
because a two-dimensional space has clear advantages in terms of parsimony, ease of interpretability, and visualization. 
As mentioned above,
it is natural to interpret $\mR^2$ as an interaction map rather than an ability map,
because the added value of the latent space model is that it captures deviations from the main effects $\alpha_j$ and $\beta_i$ of the Rasch model -- that is,
interactions of respondent $j$ and item $i$ -- and visualizes those interactions by embedding respondents along with items in $\mR^2$.

\subsubsection{Theoretical advantages}

Among the theoretical advantages of the proposed latent space model is the fact that it weakens the conditional independence assumptions of conventional IRT models,
along with the homogeneity assumptions of classic IRT models.

\paragraph{Conditional independence assumptions}

The proposed latent space model is based on the following conditional independence assumption:
\begin{equation*}
\mathbb{P}(\bm{Y} = \bm{y} \mid \th,\, \be,\, \gamma,\, \bm{A},\, \bm{B})  
\;=\; \prod_{j=1}^N\, \prod_{i=1}^I\,  \mathbb{P} (Y_{j,i} = y_{j,i} \mid \alpha_j,\, \beta_i,\, \gamma,\, \bm{a}_j,\, \bm{b}_i), 
\end{equation*}
where $\th = (\alpha_1, \dots, \alpha_N)$,
$\be = (\beta_1, \dots, \beta_I)$, 
$\bm{A} = (\bm{a}_1, \dots, \bm{a}_N)$,
and $\bm{B} = (\bm{b}_1, \dots, \bm{b}_I)$. 
In words, 
the item responses are assumed to be independent conditional on the positions of respondents and items in the latent space,
and the respondent and item attributes. 
This conditional independence assumption is weaker than the conditional independence of the Rasch model,
which requires that item responses are independent conditional on respondent and item attributes.
So the latent space model relaxes the conditional independence assumptions of the Rasch model and other classic IRT models.

The weaker conditional independence assumption of the latent space model allows for respondent-item interactions.
As a consequence, 
the latent space model can account for local dependence among item responses arising from a variety of sources,
including testlets (e.g., items similar in content), 
learning and practice effects, 
or repeated measurements, 
as well as person dependence stemming from shared school or family memberships (or even unobserved memberships). 

\paragraph{Homogeneity assumptions}

In addition, 
the latent space model drops some of the homogeneity assumptions made by conventional IRT models.
For example,
consider two respondents $j_1$ and $j_2$ with identical abilities,
who are located at distances $d(\bm{a}_{j_1},\, \bm{b}_i) < d(\bm{a}_{j_2},\, \bm{b}_i)$ from item $i$.
Then respondent $j_1$ has a higher probability of giving a correct response to item $i$ than respondent $j_2$,
despite the fact that $j_1$ and $j_2$ have identical abilities.
A similar scenario arises when two items $i_1$ and $i_2$ have identical difficulty levels and distances $d(\bm{a}_j,\, \bm{b}_{i_1}) > d(\bm{a}_j,\, \bm{b}_{i_2})$ from some respondent $j$,
which implies that item $i_1$ is less likely to be answered correctly than item $i_2$,
despite identical difficulty levels.

\begin{table}[htbp]
\centering
\caption{Hypothetical item response matrix consisting of four respondents 1, 2, 3, 4 and six items I1, I2, I3, I4, I5, I6. 
The four respondents show two response patterns. 
Respondents 1 and 2 give correct responses to Items I1--I3 only, 
whereas Respondents 3 and 4 give correct responses to Items I4--I6 only.}\s

\label{tab:homogenity_1}
\begin{tabular}{ cccc ccc}
  \hline
  Response & I1 &I2 & I3 & I4 & I5 & I6 \\ 
  \hline
  1 & 1 & 1 & 1 & 0 & 0 & 0 \\ 
  2 & 1 & 1 & 1 & 0 & 0 & 0 \\ 
  3 & 0 & 0 & 0 & 1 & 1 & 1 \\ 
  4 & 0 & 0 & 0 & 1 & 1 & 1  \\ 
   \hline
\end{tabular}
\end{table}

To give a specific example of an educational assessment where such homogeneity assumptions are violated and to demonstrate how the latent space approach captures those violations,
we provide a hypothetical item response matrix in Table \ref{tab:homogenity_1} with six math test items answered by four respondents.
Suppose the first three items are algebra items, while the last three items are geometry items. 
Assume that the algebra and geometry items have identical difficulty levels. 
Table \ref{tab:homogenity_1} shows that Persons 1 and 2 have all algebra item correct but none of the geometry items, 
whereas Persons 3 and 4 have all geometry items correct but no algebra item. 
In other words,
all respondents have three correct responses,
which is an indication that all respondents have similar abilities because the difficulties of the items are the same (by assumption).
However,
despite similar abilities of the four respondents and identical difficulty levels of the items,
it is hard to believe that the response probabilities of all respondents and all items are similar.

\begin{figure}[t]
     \centering
\begin{tabular}{cc}
(a) Two response patterns &  (b) Two response patterns, with randomness \\     
     
 \includegraphics[width=0.4\textwidth]{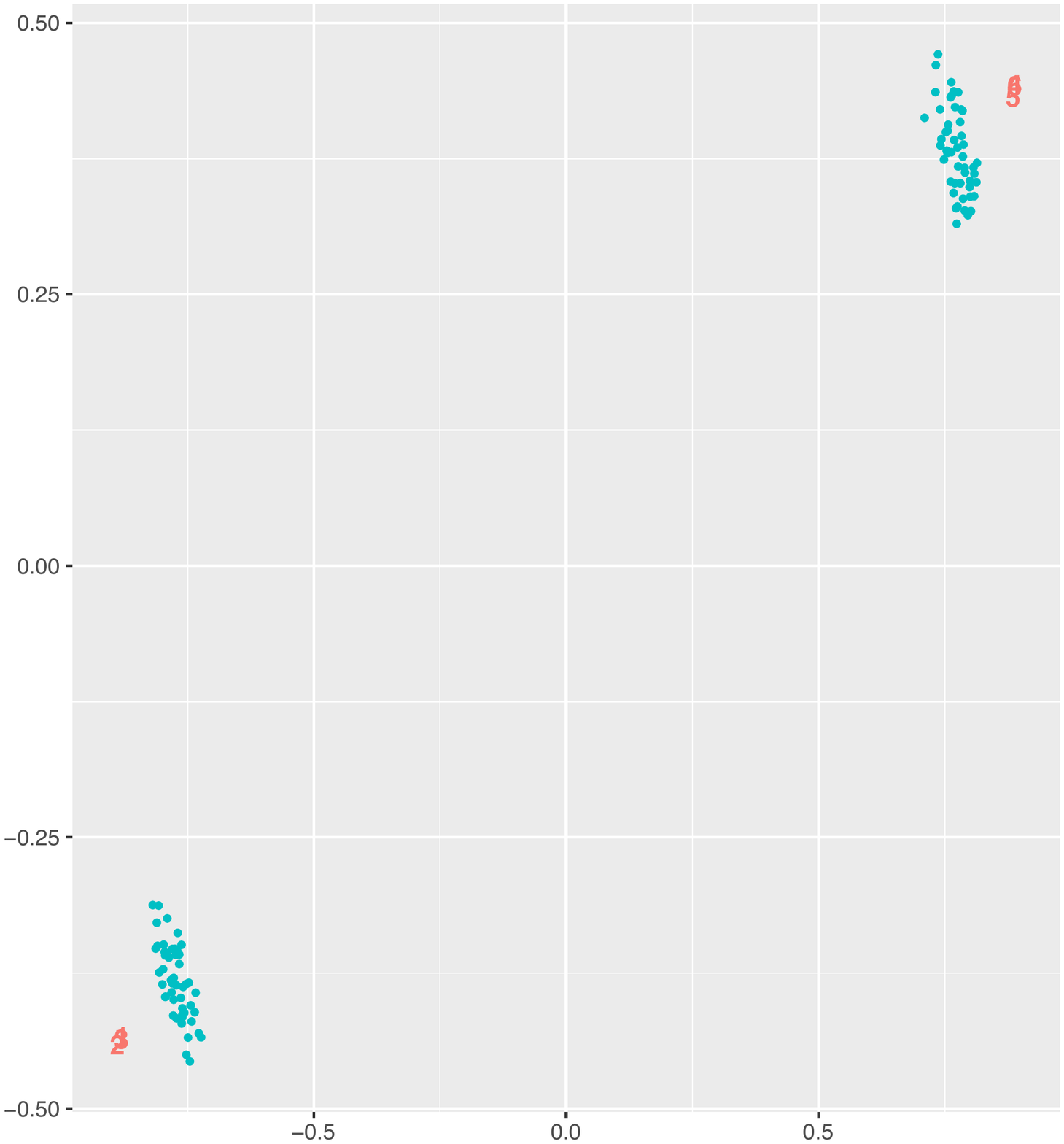} &
  \includegraphics[width=0.4\textwidth]{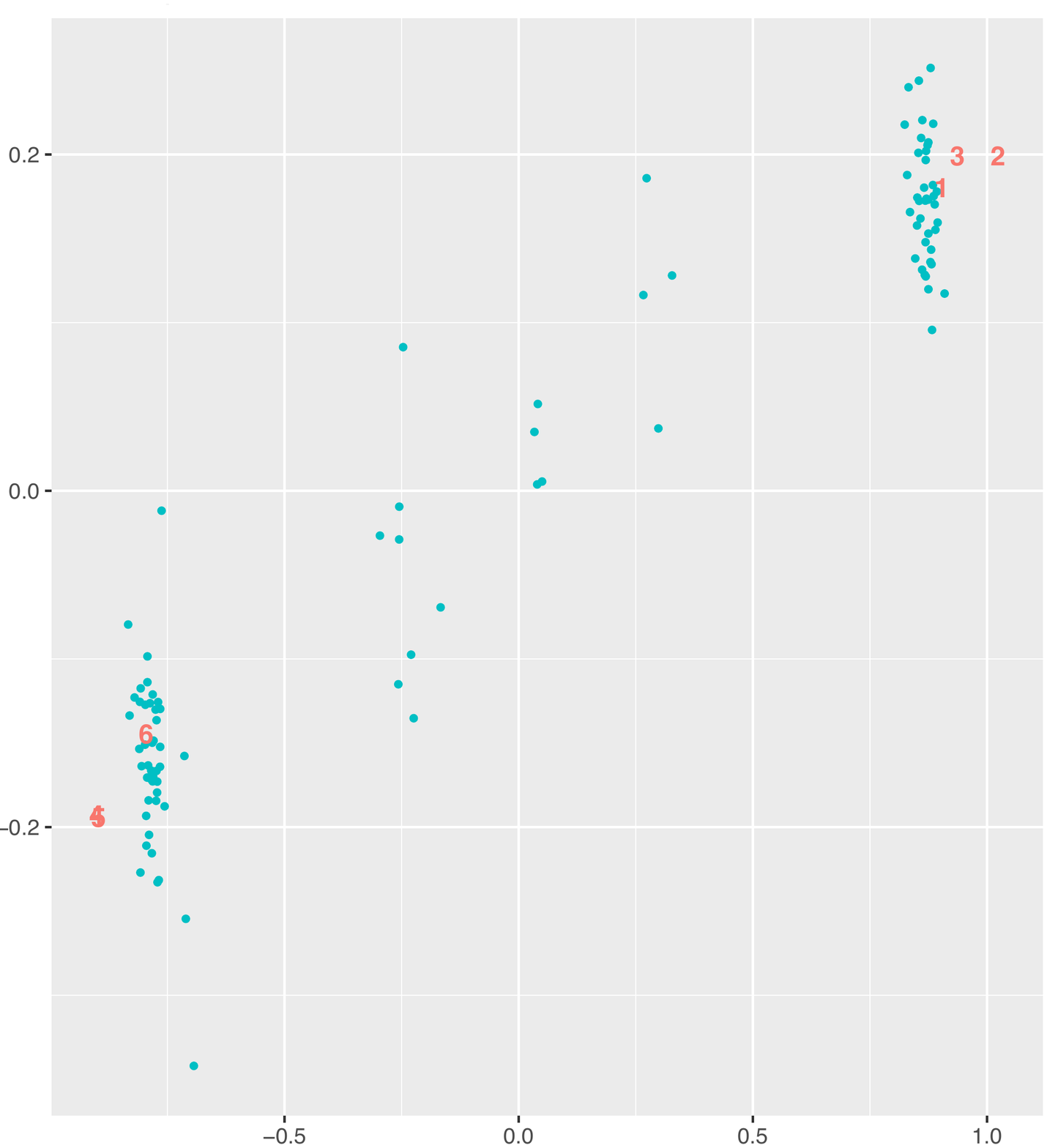} \\
 
 \end{tabular}
\caption{(a) Latent space estimated from a hypothetical data example where Respondents 1--50 (bottom left) give correct responses to Items 1--3 only, 
whereas Respondents 51--100 (top right) give correct responses to Items 4--6 only. 
$\gamma$ was estimated as 4.45 (posterior median, with 95\% posterior credible interval [3.79, 5.20]). 
(b) Latent space estimated from a hypothetical setting similar to (a):
Respondents 1--40 (bottom left) give correct responses to Items 1--3, 
while Respondents 41--80 (top right) give correct responses to Items 4--6. 
The remaining 20 respondents (middle) give random responses to Items 1--6. 
$\gamma$ was estimated as 3.62 [3.07, 4.26].
In both figures, 
blue circles represent respondents and red numbers represent items.  
}\label{fg:het}
 \end{figure}

Figure \ref{fg:het}(a) represents estimated latent space configurations based on item response data mimicking the item response matrix in Table \ref{tab:homogenity_1},
with 50 respondents giving correct responses to the first three items but incorrect responses to the last three items, 
while the other 50 respondents give incorrect responses to the first three items and correct responses to the last three items. 
Figure \ref{fg:het}(b) represents estimated latent space configurations based on item response data including random response patterns -- 80 respondents have response patterns similar to the response matrix in Table \ref{tab:homogenity_1}, 
while the other 20 respondents give random responses to Items 1--6.
Figure \ref{fg:het} reveals that the latent space approach separates the two groups of respondents in both cases, with and without randomness.
While the scenario in Figure \ref{fg:het}(a) may be an extreme-case scenario,
the scenario in Figure \ref{fg:het}(b) is more realistic, and may very well be encountered in practice.

% % attribute model
% > post.m.g 
% [1] 2.863969
% > quantile(post.g, probs=c(0.025, 0.975))
%     2.5%    97.5% 
% 2.511783 3.227793 
% > post.b <- matrix(unlist(fit$chain.keep$b), nrow=M.keep, byrow=T) # , 2, mean)
% > post.m.b <- apply(post.b, 2, mean)
% > post.m.b
% [1] 4.122898 4.088603 4.097432 4.092701 4.142447 4.116112
% > post.t <- matrix(unlist(fit$chain.keep$tau), nrow=M.keep, byrow=T) # , 2, mean)
% > post.m.t <- apply(post.t, 2, mean)
% > quantile(post.t, probs=c(0.025, 0.975))
%      2.5%     97.5% 
% 0.3519366 0.7204389 
% > post.th <- matrix(unlist(fit$chain.keep$u), nrow=M.keep, byrow=T) # , 2, mean)
% > post.m.th <- apply(post.th, 2, mean)
% > head(post.m.th)
% [1] 0.06018088 0.07661095 0.11744397 0.06239034 0.05936519 0.04634995
% > summary(post.m.th)
%   Min. 1st Qu.  Median    Mean 3rd Qu.    Max. 
% 0.01695 0.05026 0.06246 0.06232 0.07408 0.11744 
% > M.keep
% [1] 5000
% > 

% % no attribute model 
% > post.m.mu
% [1] 5.437145
% > quantile(post.mu, probs=c(0.025, 0.975))
%     2.5%    97.5% 
% 5.414378 5.478152 
% > post.m.g
% [1] 3.578272
% > quantile(post.g, probs=c(0.025, 0.975))
%     2.5%    97.5% 
% 3.280159 3.905199 
% > 

% \end{figure}

\subsection{Related models}

We review related models,
excluding those we have already reviewed in Section \ref{sec:intro}.

\subsubsection{Other models with relaxed assumptions}

As discussed above,
the proposed latent space model weakens the assumptions of classic IRT models, 
allowing for unobserved heterogeneity and dependence among item responses. 
Other approaches to relaxing those assumptions include polytomous item models, testlet and bifactor models, 
interaction effects models (e.g., \citealp{wainer:87, wilson:95}),
finite mixture models with latent classes \citep[e.g.,][]{rost:90},
and multilevel models \citep[e.g.,][]{fox:01}. 

Although these approaches have been applied successfully in applications,
they are not free of limitations. 
For instance, 
many of them require the dependence structure of items and respondents to be known prior to data analysis,
which is a strong assumption.
The latent space model does not require knowledge of the interaction structure.
In addition, 
the discussed approaches relax some but not all of the assumptions:
e.g.,
finite mixture models allow for heterogeneity between latent classes,
but assume homogeneity within latent classes.
%\textcolor{black} 
{In addition,
finite mixture models assume that there are latent classes,
which is equivalent to assuming that there is an unobserved, discrete metric space.
In contrast, 
the proposed latent space model assumes that the unobserved metric space is continuous (Euclidean) rather than discrete,
offering more flexibility to represent respondent-item interactions.}

\subsubsection{Other models with interactions among respondents and items} 

\hide{
We discussed that the proposed latent space model can be viewed as a generalization of the Rasch model with an additional distance term. 
In the sense that distances represents relations  between respondents and items, after controlling for the main effects of respondent and item attributes, distances can be regarded as interactions between respondents and items that are not explained with the main effects. 
}

\paragraph{Two-parameter IRT model}

An alternative model that captures interactions among respondents and items is the two-parameter IRT model,
which assumes that
\begin{equation}\label{eq:4}
\mbox{logit} (\mathbb{P}(Y_{j,i} = 1 \mid \alpha_j,\, \beta_i,\, \lambda_i)) \;=\; \lambda_i\; \alpha_j + \beta_i,
\end{equation}
\noindent 
where $\lambda_i \in \mR$.
The term $\lambda_i\; \alpha_j$ captures interactions of item $i$ and respondent $j$.
The latent space approach has an important advantage over the two-parameter IRT model:
It embeds both respondents and items in a low-dimensional space,
helping visualize interactions of respondents and items.

\paragraph{Interaction IRT model}

A more general interaction model assumes that
\begin{equation}\label{eq:3}
\mbox{logit} (\mathbb{P}(Y_{j,i} = 1 \mid \alpha_j,\, \beta_i,\, \epsilon_{j,i}) ) \;=\; \alpha_j + \beta_i + \epsilon_{j,i}, \end{equation}
\noindent 
where $\epsilon_{j,i} \in \mR$ represents the interaction of respondent $j$ and item $i$.
The latent space model can be viewed as a special case of the interaction model,
corresponding to
\[
\epsilon_{j,i}
\;=\; - \gamma\; d(\bm{a}_j,\, \bm{b}_i).
\]
In other words,
the latent space model makes the implicit assumption that the interaction effects $\epsilon_{j,i}$ are of the form
$\epsilon_{j,i} = - \gamma\; d(\bm{a}_j,\, \bm{b}_i)$,
where the distances $d(\bm{a}_j,\, \bm{b}_i)$ satisfy reflexivity, symmetry, and the triangle inequality,
as described in Section \ref{development}.
While the proposed latent space model is a special case of the interaction model,
it has three advantages over the interaction model.
First, 
the latent space model can be estimated,
whereas the interaction model cannot be estimated unless additional parameter constraints are imposed, 
because in practice we have a single observation (i.e., item response) for each pair of respondents and items.
Second,
the latent space model captures transitivity in item response data thanks to the triangle inequality:
e.g.,
if the positions of two respondents $j_1$ and $j_2$ are close to the position of item $i$ in the latent space,
then the two respondents are fairly close to each other,
by the triangle inequality;
likewise,
if two items $i_1$ and $i_2$ are close to respondent $j$,
then the two items are fairly close to each other.
The assumption that item response data are transitive makes sense in applications.
Therefore,
while the latent space model is more restrictive than the general interaction model,
the restrictions make sense in practice,
and facilitate estimation.
Last,
but not least,
the latent space approach provides an interaction map of respondents and items.

\paragraph{Bilinear mixed effects models and related models}
The multiplicative effects version of the latent space model is related to the bilinear mixed effects model of \citet{hoff:05}, 
the additive and multiplicative effects models of \citet{Ho18},
and the latent factor models of \citet{agarwal:09}.
For example,
the bilinear mixed effects models of \citet{hoff:05} are models of network data,
such as friendships among $N$ students.
Bilinear mixed effects models add a multiplicative effect of the form  $\bm{a}_j^\top\, \bm{b}_i$ to the log odds of a friendship between students $i$ and $j$.
The multiplicative effects version of the proposed latent space models resembles the multiplicative effects in the above-mentioned models,
but multiplicative effects are more difficult to interpret,
as mentioned at the beginning of Section \ref{sec:model}. 

\paragraph{Differential item functioning}

Last,
but not least,
IRT models for studying differential item functioning (DIF) can be seen as special cases of interaction models, 
where an interaction term is formed with a known categorical attribute of respondents (e.g., gender) and an item indicator. 
Conventional DIF models, however, require pre-knowledge of the respondent attribute.
The proposed latent space model does not require such pre-knowledge.

\section{Bayesian inference}\label{sec:estimation}

\subsection{Markov chain Monte Carlo (MCMC)}\label{sec:mcmc}

We propose a fully Bayesian approach for estimating the proposed latent space model,
using MCMC methods.
Bayesian inference is preferable to maximum likelihood due to under-identification of the latent space positions.

We use the following priors:
\[
\begin{array}{llllllllll}
\alpha_j \mid \sigma^2 &\ind& \mbox{N}\left(0,\, \sigma^2\right),\;\;\;\; \sigma^2 > 0,\;\;\;\; j = 1, \dots, N\s
\\
\beta_i \mid \tau^2_\beta &\ind& \mbox{N}\left(0,\, \tau^2_\beta\right),\;\;\;\; \tau_\beta^2 > 0,\;\;\;\; i = 1, \dots, I\s
\\
\log \gamma \mid \mu_\gamma,\, \tau^2_\gamma &\sim&  \mbox{N}\left(\mu_\gamma,\, \tau^2_\gamma\right),\;\;\;\; \mu_\gamma \in \mR,\;\;\;\; \tau^2_\gamma > 0\s
\\
\sigma^2 \mid a_\sigma,\, b_\sigma &\sim& \mbox{Inv-Gamma} \left(a_\sigma,\, b_\sigma\right),\;\;\;\; a_\sigma > 0,\;\;\;\; b_\sigma > 0\s
\\
\bm{a}_j &\iid& \mbox{MVN}_p\left(\bm{0},\, \bm{I}_p\right),\;\;\;\; j = 1, \dots, N\s
\\
\bm{b}_i &\iid& \mbox{MVN}_p\left(\bm{0},\, \bm{I}_p\right),\;\;\;\; i = 1, \dots, I,
\end{array}
\]
where $\bm{0}$ is a $p$-vector of zeroes and $\bm{I}_p$ is the $p \times p$ identity matrix.
%\textcolor{black}{
In principle, 
it is  possible to specify priors of distances directly rather than indirectly (i.e., by specifying priors of positions).
However, specifying a prior for distances is more challenging than specifying a prior for positions, because the distances must satisfy the triangle inequality.
As a consequence,
it is conventional in the latent space model literature to place a prior on positions rather than distances,
and we follow here convention.
%}
We use
$\tau^2_\beta = 4$,  $a_\sigma =1$,  $b_\sigma =1$, $\mu_\gamma =0.5$, $\tau^2_\gamma = 1$, 
which are not uncommon in the literature \citep[see, e.g.,][]{furr:16}. 
While these priors may seem strong,
note that the effective parameter space of many models for binary item response data is small:
e.g.,
when item responses are independent Bernoulli$(\pi)$ random variables with success probability $\pi \in (0, 1)$ and log odds $\alpha=\logit(\pi) \in \mR$,
then values of $\alpha$ outside of the interval $[-5, +5]$ correspond to probabilities close to $0$ or $1$, which are unrealistic.
Therefore, 
while the theoretical parameter space of $\alpha$ is $\mR$,
the effective parameter space is a subset of $\mR$, 
e.g., 
$[-5, +5]$.
As a consequence,
using priors that place most probability mass on $[-5, +5]$ are reasonable,
and so are the priors suggested above.
The priors described above are used throughout the remainder of the paper,
unless stated otherwise.

The posterior of the parameters $\bm\alpha$, $\bm\beta$, and $\gamma$ and the unobserved positions of respondents $\bm{A}$ and items $\bm{B}$,
given an observation $\bm{y}$ of $\bm{Y}$,
is proportional to
\begin{equation}
\begin{array}{lllll}
f (\bm{\alpha},\; \bm{\beta},\; \gamma,\; \bm{A},\; \bm{B} \mid \bm{y}) 
&\propto& \left[\displaystyle\prod_{j=1}^N f (\alpha_j)\right]\; \left[\displaystyle\prod_{i=1}^I f(\beta_i)\right]\; f(\gamma)\; \left[\displaystyle\prod_{j=1}^N  f(\bm{a}_j)\right]\; \left[\displaystyle\prod_{i=1}^I f(\bm{b}_i)\right]\s\s
\\
&\times& \left[\displaystyle\prod_{j=1}^N \displaystyle\prod_{i=1}^I \mathbb{P}\Big(Y_{j,i}=y_{j,i} \mid \alpha_j,\; \beta_i,\; \gamma,\; \bm{a}_j,\; \bm{b}_i\Big)\right],
\end{array}
\end{equation}
where,
in an abuse of notation,
we use $f(.)$ to denote the prior and posterior probability density functions of the parameters as well as the positions of respondents and items.

We sample from the posterior by using an MCMC algorithm that updates the parameters and the positions of respondents and items at iteration $t$ as follows:
\begin{enumerate}
    \item Propose $\alpha_j^\star$ from a symmetric proposal distribution and accept the proposal with probability
    \[
    %r_\alpha \Big(\alpha_j^\star,\, \alpha_j^{(t)}\Big) = 
\min\left(1,\; \frac{f\Big(\alpha_j^\star \mid \yy,\, \bm{A},\, \bm{B},\, \be,\, \gamma\Big)} {f\Big(\alpha_j^{(t)} \mid \yy,\, \bm{A},\, \bm{B},\, \be,\, \gamma\Big)}\right).
\]
    \item Propose $\beta_i^\star$ from a symmetric proposal distribution and accept the proposal with probability 
    \[ 
%    r_\beta \Big(\beta_i^\star,\, \beta_i^{(t)} \Big) = 
\min\left(1,\; \frac{f\Big(\beta_i^\star \mid \yy,\, \bm{A},\, \bm{B},\, \th,\, \gamma\Big)} {f\Big(\beta_i^{(t)} \mid \yy,\, \bm{A},\, \bm{B},\, \th,\, \gamma\Big)}\right). 
   \]
    \item Propose $\gamma^\star$ from a symmetric proposal distribution and accept the proposal with probability
    \[
%    r_\gamma\Big(\gamma^\star,\, \gamma^{(t)}\Big) = 
\min\left(1,\; \frac{f\Big(\gamma^\star \mid \yy,\, \bm{A},\, \bm{B},\, \th,\, \be\Big)}{f\Big(\gamma^{(t)} \mid \yy,\, \bm{A},\, \bm{B},\, \th,\, \be\Big)}\right).
    \]
    \item Propose $\bm{a}_j^\star$ from a symmetric proposal distribution and accept the proposal with probability
    \[
%    r_{\bm{a}_j} \Big(\bm{a}_j^\star,\, \bm{a}_j^{(t)}\Big) = 
\min\left(1,\; \frac{f\Big(\bm{a}_j^\star \mid \yy,\, \bm{A}_{-j},\, \bm{B},\, \th,\, \be,\, \gamma\Big)}{f\Big(\bm{a}_j^{(t)} \mid \yy,\, \bm{A}_{-j},\, \bm{B},\, \th,\, \be,\, \gamma\Big)}\right),
    \]
where $\bm{A}_{-j} = (\bm{a}_1, \dots, \bm{a}_{j-1}, \bm{a}_{j+1}, \dots, \bm{a}_N)$.
    \item Propose $\bm{b}_i^\star$ from a symmetric proposal distribution and accept the proposal with probability
    \[
%    r_{\bm{b}_i} \Big(\bm{b}_i^\star,\, \bm{b}_i^{(t)}\Big) = 
\min\left(1,\; \frac{f\Big(\bm{b}_i^\star \mid \yy,\, \bm{A},\, \bm{B}_{-i},\, \th,\, \be,\, \gamma\Big)} 
    {f\Big(\bm{b}_i^{(t)} \mid \yy,\, \bm{A},\, \bm{B}_{-i},\, \th,\, \be,\, \gamma\Big)}\right),
    \]
where $\bm{B}_{-i} = (\bm{b}_1, \dots, \bm{b}_{i-1}, \bm{b}_{i+1}, \dots, \bm{b}_I)$.
    \item Sample $\sigma^2$ from its full conditional distribution:
    \[
    \sigma^2 \sim \mbox{Inv-Gamma}\left(a_\sigma + \frac{N}{2},\; b_\sigma + \frac{\sum_{j=1}^N \alpha_j^2}{2}\right).
\]    
\end{enumerate}
As symmetric proposal distributions,
we use (multivariate) Gaussian distributions centered at the current values of the parameters or the positions of respondents and items,
with diagonal variance-covariance matrices.
\textcolor{black}{The variances of the (multivariate) Gaussians are set to achieve a good performance of the algorithm (with an acceptance rate of 0.3)}. 
To detect non-convergence of the MCMC algorithm,
we use trace plots along with the Gelman-Rubin diagnostic \citep{gelman:92}.
\textcolor{black} {The MCMC algorithm was written in {\tt R.} 
The {\tt R} code, 
along with an example dataset, 
can be found in the supplementary materials. 
}

\subsection{Identifiability}

The log odds of a correct response is invariant to translations, reflections, and rotations of the positions of respondents and items,
because the log odds depends on the positions through the distances,
and the distances are invariant under the said transformations.
As a consequence,
the likelihood function is invariant under the same transformations.
%to reflections, rotations, and translations of the positions of items and respondents.
The same form of identifiability issue arises in latent space models of network data \citep{Hoff:2002}.
Such identifiability issues can be resolved by post-processing the MCMC output with Procrustes matching \citep{Gower:1975}. 
%\textcolor{black}{Bayesian estimation is preferable to maximum likelihood estimation for models that involve latent spaces due to the technical difficulties stemming from  under-identification of the latent space positions.} 
%which is the method adopted by \citet{Hoff:2002} and \citet{Handcock:2007}. For implementation, we  first identify the MCMC iteration that achieves the highest likelihood; then select a collection of the latent positions  from the iteration  as the reference points; lastly, apply Procrustes transformation to the posterior draw of the latent positions at each iteration. For instance, suppose $\bm{z}_0$ is the $p$-dimensional matrix of the reference positions of respondents and $\bm{z}_t$ is the matrix of the positions estimated at iteration $t$. The matched position matrix $\bm{z}_t^*$ for $\bm{z}_t$ is obtained based on the following transformation: 
%\[
%%\bm{z}_t^* =  s \cdot \bm{z}_t \cdot \RR + 1\bm{v}'
%\]
%\noindent where $s$ is the scale parameter, $\RR$ is the rotation matrix, and $\bm{v}$ is the translation vector which are chosen to approximate $\bm{z}_0$ \citep{borg:97}. % (Borg and Groenen, 1997). 
However,
the results need to be interpreted with care, 
because there are many latent space configurations that give rise to the same distances.
So an estimated latent space should be interpreted in terms of the relative distances between positions, rather than the actual positions.

\hide{

{\color{black}
\subsection{Model Misspecification}

The model fit and/or the accuracy of local  dependence captured  with the proposed model may  be affected  by mis-specified  prior  distributions.   One  can  check, for instance,  if priors that allow  for  multiple  manifest  or  latent  groups  for $a$ and $b$ give a  better  fit   and/or  a  better  explanation  of  the  local  dependence in the data. It is also possible that a different choice on metric space, the form of the distance function, or the latent space dimension may result in a different model fit. One may want to conduct a sensitivity analysis to evaluate the impacts of these and other factors on model fit and accuracy. 
}

\textcolor{red}{mj: Ick Hoon worries that this paragraph may invite reviewers to ask us to do the simulation study.  }

}

{\color{black}
\subsection{Model Selection} \label{sec:model_selection}

In practice,
given a dataset,
it is natural to ask:
Did the Rasch model with $\gamma=0$ or the latent space model with $\gamma > 0$ generate the data?
If the latent space model generated the data,
it is appropriate to base conclusions regarding respondents and items on the latent space model,
including the interaction map provided by the latent space model.
Otherwise the Rasch model suffices.

To determine whether the Rasch model with $\gamma=0$ or the latent space model with $\gamma > 0$ generated the data,
we use a model selection approach based on spike-and-slab priors \citep{Ishwaran:05}.
We specify a spike-and-slab prior for $\log \gamma$ by specifying a prior consisting of two component distributions:
a spike prior $N_{spike}(\mu_{\gamma_0},\, \tau^2_{\gamma_0})$ with a small variance $\tau^2_{\gamma_0} > 0$ that places most of its probability mass in a small neighborhood of $0$,
and a slab prior $N_{slab}(\mu_{\gamma_1},\, \tau^2_{\gamma_1})$ with a large variance $\tau^2_{\gamma_1} > 0$ that distributes its probability mass across the parameter space.
In other words,
the prior of $\log \gamma$ may be expressed as 
\[
\log \gamma\; \sim\; (1-\delta)\; N_{spike}(\mu_{\gamma_0},\,  \tau^2_{\gamma_0}) + \delta\; N_{slab}(\mu_{\gamma_1},\,  \tau_{\gamma_1}^2),
\]
where $\delta \in \{0, 1\}$.
If the posterior probability of the event $\delta=0$ is less than $0.5$,
we choose the Rasch model with $\gamma=0$,
otherwise we choose the latent space model with $\gamma > 0$.
The posterior probability can be approximated by the proportion of times we observe the event $\delta=1$ in a Markov chain Monte Carlo sample from the posterior. 
We choose as a prior for $\omega = p(\delta = 1 \mid \omega) \in [0, 1]$ the $\mbox{Beta}(1,\, 1)$ distribution.
As a spike prior,
we use $N_{spike}(-3,\,  1)$,
so that the distribution of $\log \gamma \mid \delta=0$ has mode 
$0.02$,
mean $0.08$, 
and standard deviation $0.01$.
As a slab prior,
we use $N_{spike}(0.5,\, 1)$,
with mode $.0.61$,
mean $2.72$,
and standard deviation $3.56$.
As pointed out in Section \ref{sec:mcmc},
the effective parameter space of many models for binary item response data is small,
so the slab prior is not unreasonable.

The model selection method described is applied to the two real data applications provided in Section \ref{sec:application}. 
We evaluate the accuracy of the model selection method via simulations in Section \ref{sec:simulation}. 
}

%> mode <- function(mu, sigma2) { exp(mu - sigma2) }
%> mean <- function(mu, sigma2) { exp(mu + sigma2 / 2) }
%> sd <- function(mu, sigma2) { sqrt((exp(sigma2) - 1) * exp(2 * mu + sigma2)) }
%> mode(-3, 1)
%[1] 0.01831564
%> mean(-3, 1)
%[1] 0.082085
%> sd(-3, 1)
%[1] 0.1075997
%> mode(1/2, 1)
%[1] 0.6065307
%> mean(1/2, 1)
%[1] 2.718282
%> sd(1/2, 1)
%[1] 3.563212
%the uninformative prior for $\log \gamma$ described  in Section \ref{sec:mcmc}  
}

\section{Applications}\label{sec:application}

To demonstrate the latent space approach, 
we provide two empirical examples.

\subsection{Example 1: Attitudes to Abortion}

\subsubsection{Data and Estimation}

As a first example, we used the attitudes-to-abortion scale that came from \citet{Planning:1987}. 
Seven items were included in the scale, 
which ask respondents whether abortion should be legal in each of the following seven scenarios: 
\begin{enumerate}
    \item The woman decides on her own that she does not wish to have the child 
    \item The couple agree that they do not wish to have the child
    \item The woman is not married and does not wish to marry the man
    \item The couple cannot afford any more children. 
    \item There is a strong chance of a defect in the baby
    \item The woman's health is seriously endangered by the pregnancy 
    \item The woman became pregnant as a result of rape
\end{enumerate}

%The data include binary responses from 642 respondents, collected over the four panel waves.  
Binary responses to the seven items were collected, where 
response `Yes' was coded as 1 and response `No'  was coded as 0. 
The mean proportion of `Yes' was 0.42, 0.52, 0.47, and 0.53 for Items 1 to 4, while 0.86, 0.94, and 0.93 for Items 5 to 7,
respectively  ($N$=642).\footnote{\textcolor{black}{Different subsets or versions of the data have been used in the literature. 
We used the data pre-processed by \citet{skrondal:04}, 
which include responses from 734 respondents. 
We analyzed the version after deleting respondents with no item responses assuming missing at random \citep{skrondal:04}.}} 
The positive response proportion was quite high for the last three items that describe rather extreme situations in which  most respondents are likely to endorse. 
To implement MCMC, we specified the priors as we described in Section \ref{sec:estimation}. 
For $\beta$, we chose a stronger prior with $\tau^2_\beta = 1$ because otherwise the MCMC did not converge well due to the boundary effects of probability for Items 5-7 (the positive answer probability was too close to 1). 
%For the model with attribute effects, 
\textcolor{black}{We selected the tuning parameters (standard deviations of the proposal distributions) to ensure a reasonable acceptance rate as follows: 2.2 for $\alpha_j$, 0.5 for $\beta_i$, 0.1 for $\gamma$, 1.7 for $\bm{a}_j$, and 0.4 for $\bm{b}_i$.} 
The MCMC run included 20,000 iterations with the first 10,000 iterations discarded as a burn-in period. The computation took approximately \textcolor{black}{28 minutes}  for the latent space model on a standard computer. Trace plots showed reasonable convergence of the sampler (convergence evidence was provided  in  Appendix B of the supplement). 
\textcolor{black}{In addition,
we used the Gelman-Rubin diagnostic \citep{gelman:92} to detect possible non-convergence.
We ran the model with three sets of random starting values; the scale reduction factor  was smaller than 1.06 for all model parameters, 
suggesting that there are no signs of non-convergence.
}
%> post.m.pi
%[1] 0.9937
\textcolor{black}{We implemented the model selection method with the spike-and-slab prior, described in Section \ref{sec:model_selection}. % to evaluate the relevance of applying the proposed latent space model over the Rasch model.  
The posterior inclusion probability of $\delta$ was .99, in favor of the proposed model to the Rasch model. 
Hence, we move forward with the latent space item response model for the current application. 
}

%, with 342 females and 300 males included. 

%  0   1 
%300 342 
%[male and female information]

\subsubsection{Results}

\begin{figure}[htbp]
\centering
\includegraphics[width=.4\textwidth]{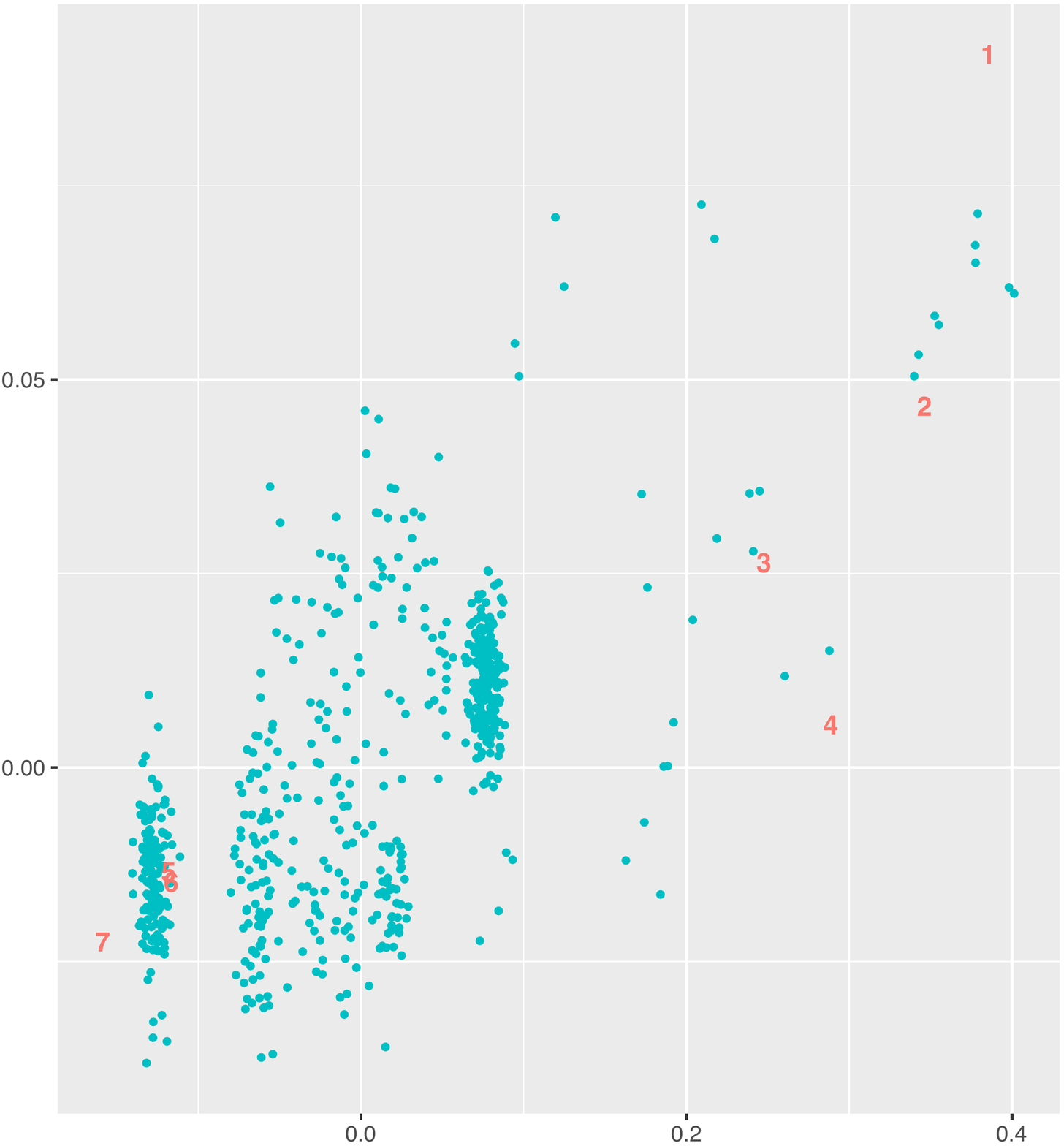} 
\caption{Estimated latent space for the attitudes to abortion data. Red numbers represent items and blue dots represent respondents.   %Items 1-4 are apart from Items 5-7. In (a), respondents appear positioned as small clusters. 
}\label{fg:zw_abortion}
\end{figure}

\paragraph{Interpreting latent space results}

Figure \ref{fg:zw_abortion} displays the estimated latent space. This latent space shows the point estimates (posterior means) of the positions but not their uncertainty, for the ease of visualization.  Uncertainty of the estimated positions, measured with the 95\% posterior credible intervals, is reported in  Appendix C of the supplement. 
The $\gamma$ parameter was estimated as 1.25 (posterior median, with 95\% posterior credible interval [0.92, 1.54]) and $\sigma$ as 2.34 (posterior median, with 95\% posterior credible interval [2.07, 2.62]). 

Roughly two item groups appear in the latent space; one group with three items in the bottom left of the space (Items 5--7) and the other group with four items in the top right side of the space (Items 1--4). Also, two respondent groups appear, with a larger group in the left bottom part of the space (near Items 5--7) and a much smaller, scattered group on the right upper side of the space (near Items 1--4). 
% Respondents in the bigger cluster, Person cluster 1, are closely located to Item cluster 1, meaning that they tended to give a positive response to the extreme items. Figure \ref{b_abortion}(a) shows that Items 5-7 are in fact very easy items compared to Items 1-4. As another confirmation, we found that about 15.7\% of the respondents who are located in the region of $Y >0.25$ (closest to Items 5-7) all said YES to Items 5 to 7 and said NO to Items 1 to 4 (except one person who gave YES to Item 3). 
%Items 5-7 involve rather extreme situations, whereas  Items 1-4 address relatively mild situations. It make sense that Items 5-7 are much easier than Items 1-4 %meaning that most people are likely to choose YES for those items that describe extreme situations 
%(see Figure \ref{compare_abortion}(a)). 
Respondents close to Items 1--4 but apart from Items 5--7  tend to give positive responses to the mild items (Items 1--4) but negative responses to the extreme items (Items 5--7). Table \ref{small_abortion} shows that the respondents in the region of $X > 0.3$ and $Y > 0.025$ (close to Items 1-4) indeed tend to choose YES to Items 1 to 4 but NO to Items 5 to 7. %These people with somemake a minority group given that their response patterns are somewhat unusual. 
Respondents close to Items 5--7  tend to give positive responses to the extreme items  but negative responses to the mild situations (Items 1--4). %These (conservative) people make a majority group in this dataset. 

\begin{table}[htbp]
\centering
\caption{Response patterns to Items 1 to 7 (I1 to I7) for respondents (ID) in the smaller person cluster, located in the bottom left corner of the latent space (which is the region of $X > 0.3$ and $Y > 0.025$).  These people tend to give positive responses to I1--I4, but negative responses to I5--I7. 
}\label{small_abortion}
\begin{tabular}{rrrrrrrr}
  \hline
  ID & I1 & I2 & I3 & I4 & I5 & I6 & I7 \\ 
  \hline
27 &   1 &   0 &   0 &   1 &   0 &   0 &   0 \\ 
  92 &   1 &   1 &   1 &   1 &   1 &   0 &   0 \\ 
  132 &   1 &   1 &   0 &   1 &   0 &   0 &   0 \\ 
  191 &   1 &   1 &   1 &   1 &   0 &   1 &   0 \\ 
  273 &   1 &   1 &   0 &   0 &   0 &   0 &   0 \\ 
  330 &   1 &   0 &   0 &   1 &   0 &   0 &   0 \\ 
  653 &   1 &   1 &   0 &   1 &   0 &   0 &   0 \\ 
  662 &   1 &   1 &   1 &   0 &   0 &   0 &   0 \\ 
  675 &   1 &   1 &   1 &   0 &   0 &   0 &   0 \\ 
%   25 &  1 &  0 &  0 &  1 &  0 &  0 &  0 \\ 
%   81 &  1 &  1 &  1 &  1 &  1 &  0 &  0 \\ 
%   111 &  1 &  1 &  0 &  1 &  0 &  0 &  0 \\ 
%   164 &  1 &  1 &  1 &  1 &  0 &  1 &  0 \\ 
%   234 &  1 &  1 &  0 &  0 &  0 &  0 &  0 \\ 
%   279 &  1 &  0 &  0 &  1 &  0 &  0 &  0 \\ 
%   292 &  0 &  1 &  1 &  1 &  0 &  1 &  0 \\ 
%   545 &  1 &  1 &  0 &  1 &  0 &  0 &  0 \\ 
%   551 &  1 &  1 &  1 &  0 &  0 &  0 &  0 \\ 
%   560 &  1 &  1 &  1 &  1 &  0 &  1 &  1 \\ 
%   564 &  1 &  1 &  1 &  0 &  0 &  0 &  0 \\ 
   \hline
\end{tabular}
\end{table}

\paragraph{Comparison with the Rasch model}

\begin{figure}[p]
\centering
\begin{tabular}{cc}
(a) LS $\beta_i$ & (b)  Rasch $\beta_i$\\
\includegraphics[width=0.45\textwidth]{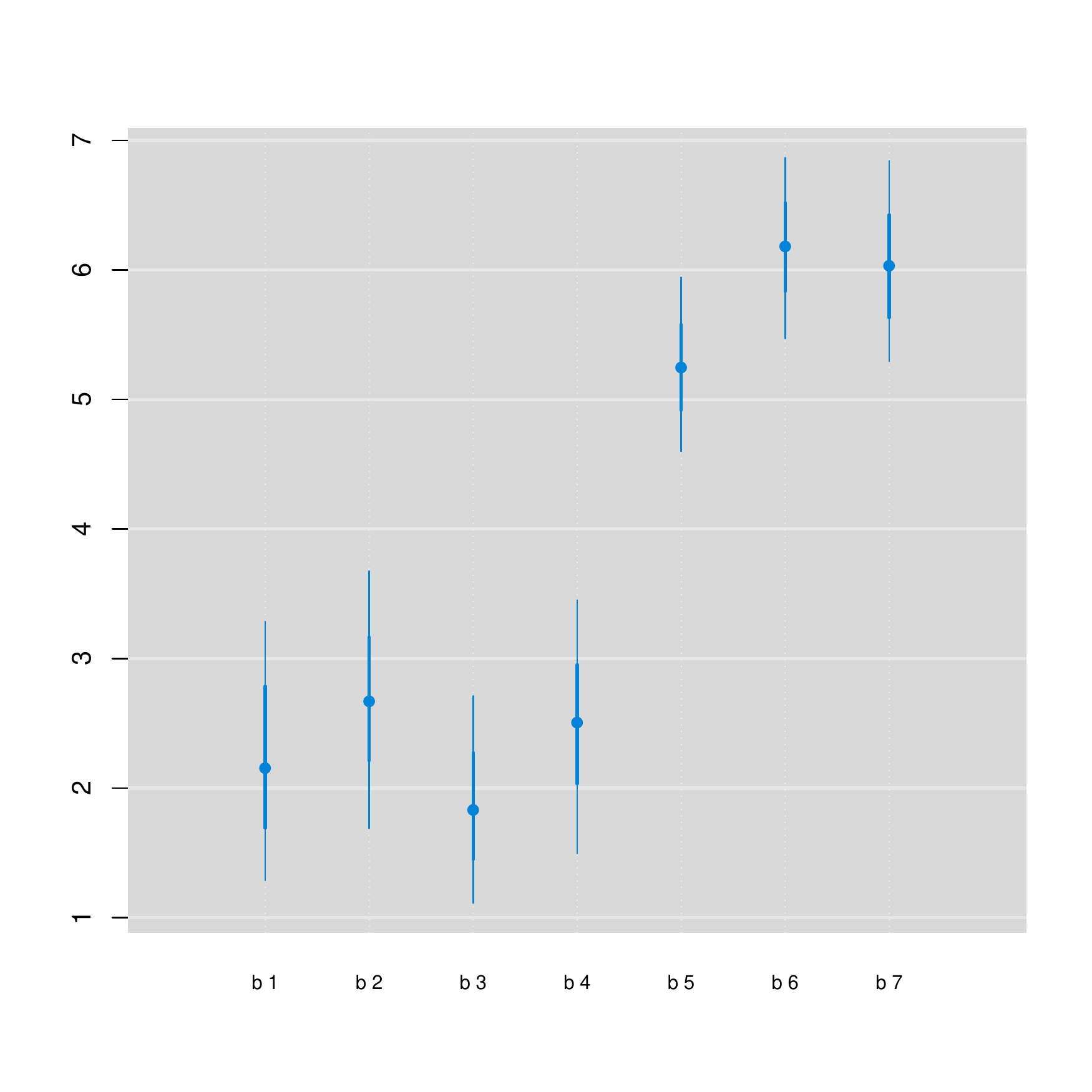} & 
\includegraphics[width=0.44\textwidth]{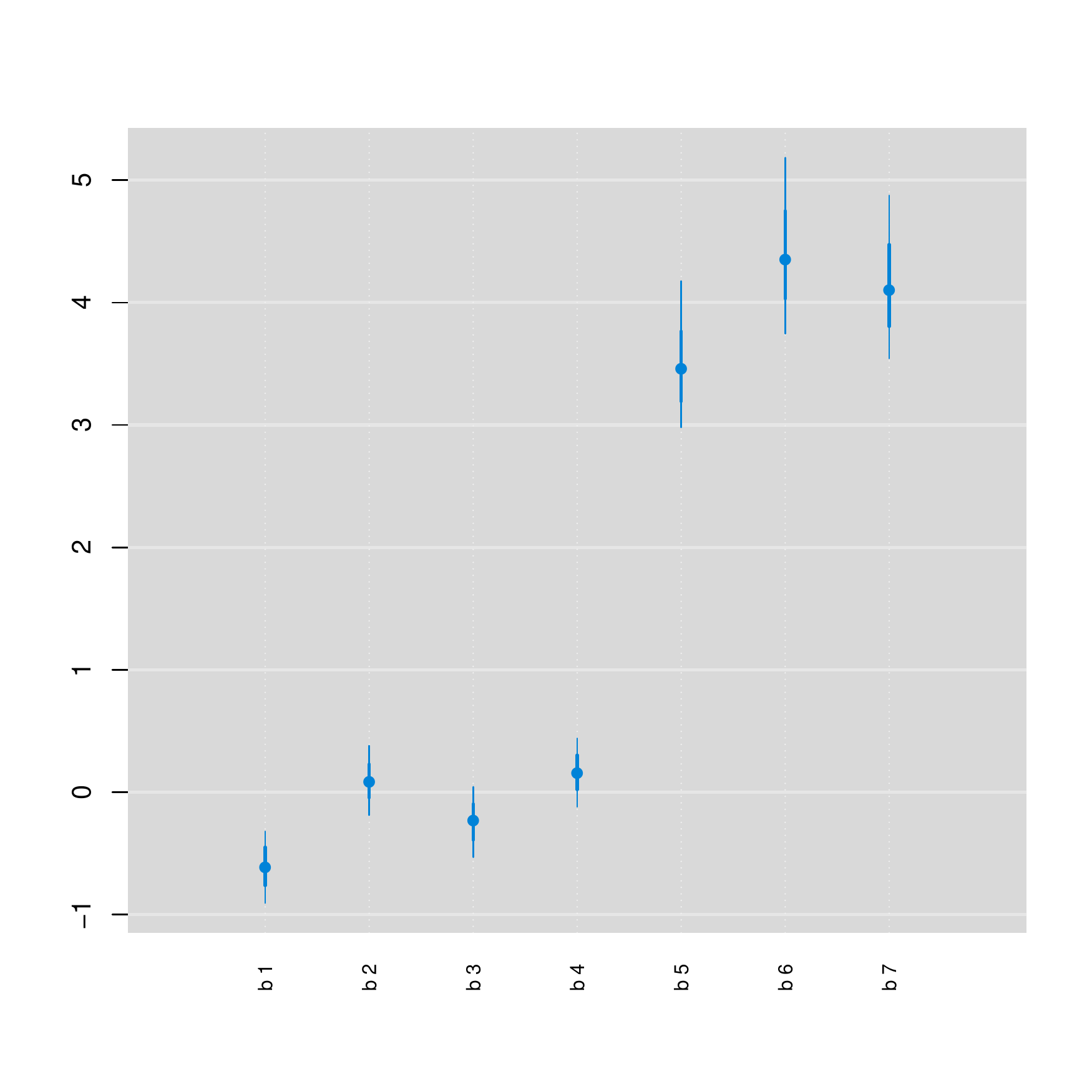}\\

(c) LS $\alpha_j$  & (d) Rasch $\alpha_j$ \\
\includegraphics[width=0.4\textwidth]{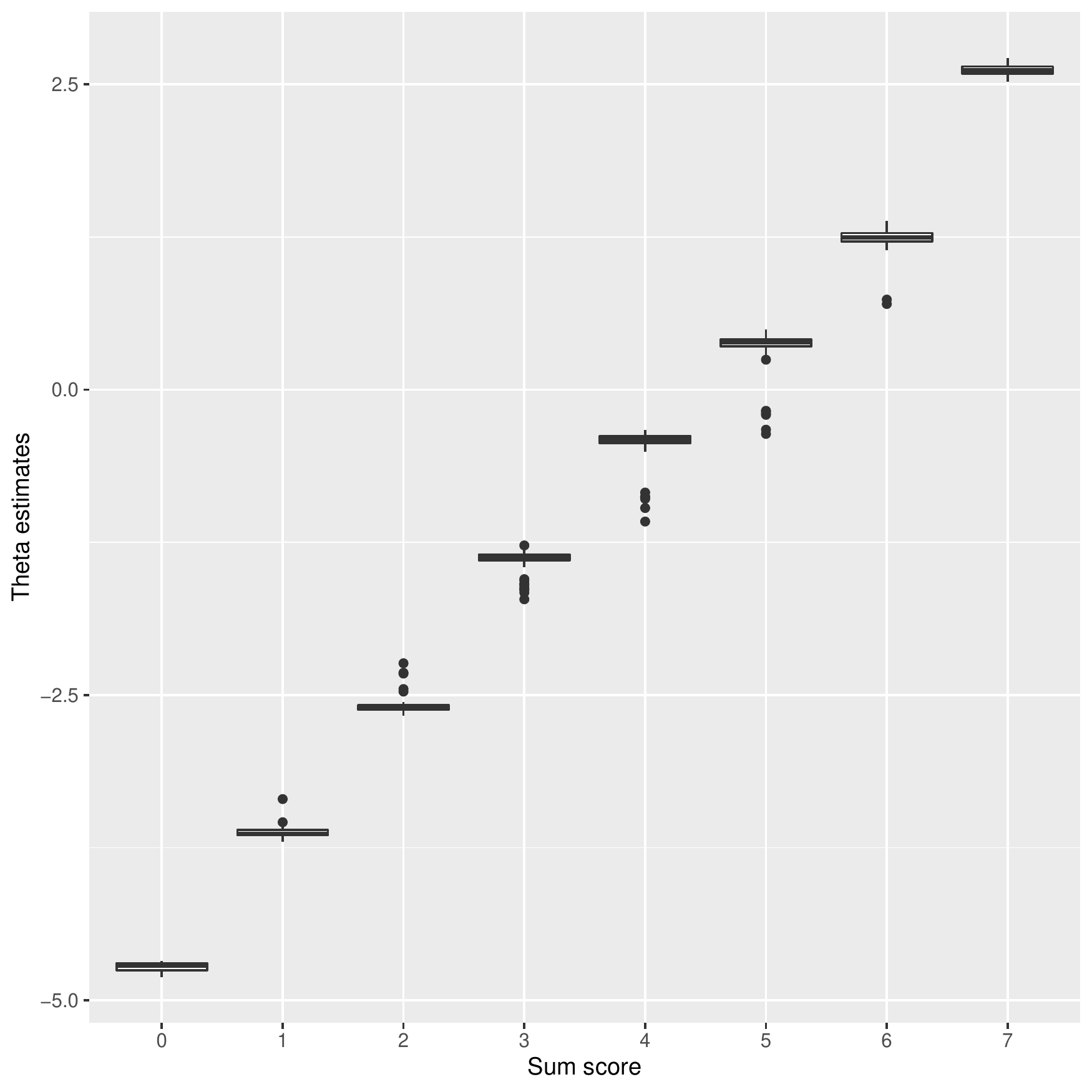} & 
\includegraphics[width=0.4\textwidth]{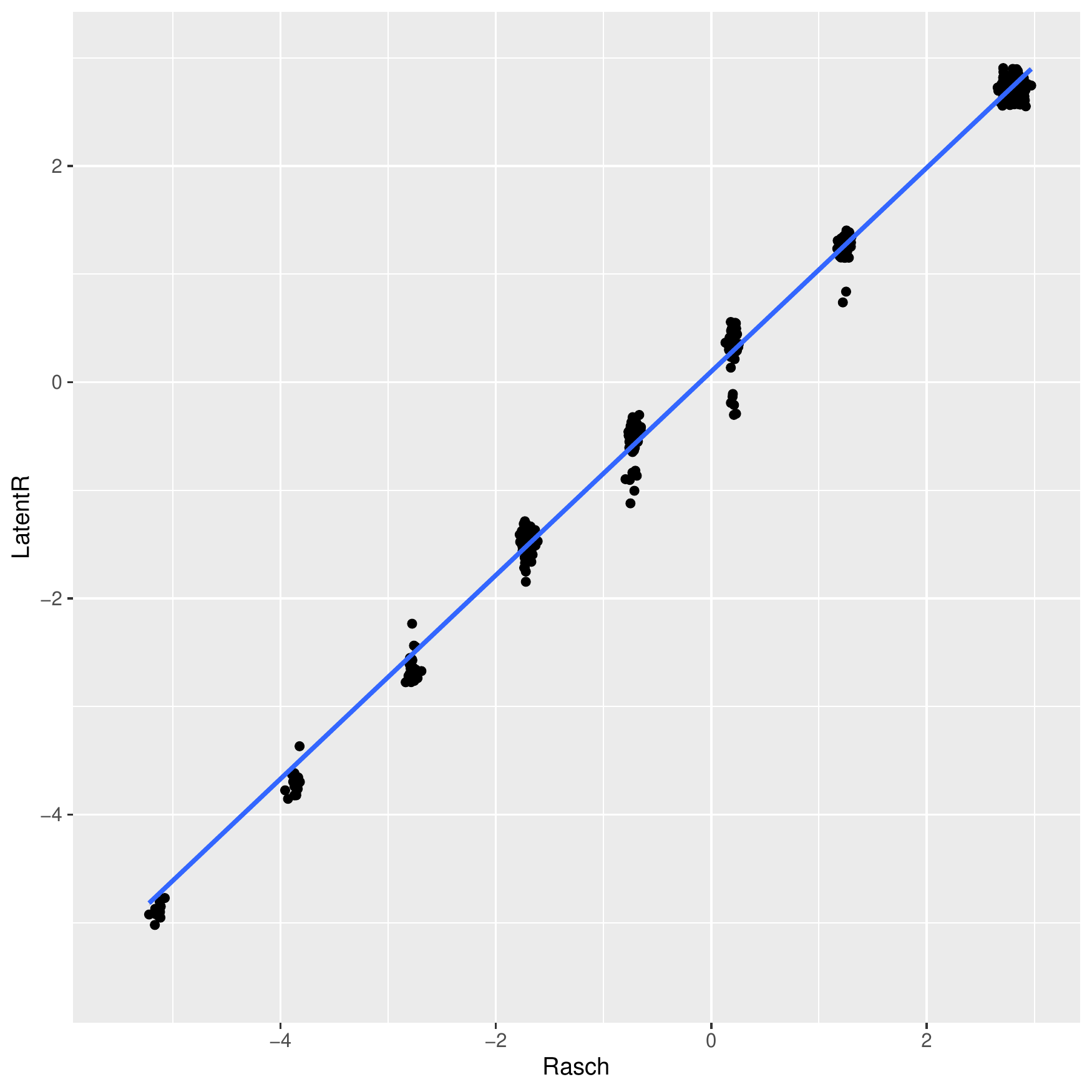} \\
\end{tabular}
\caption{\label{compare_abortion}
(a) and (b) show the 95\% posterior credible intervals for $\beta$ for Items 1 to 7 (indicated by b1 to b7) from the latent space model (LS) and the Rasch model for the attitudes to abortion data. 
(c) shows the box plot of $\alpha$ estimates per total sum score on the X-axis (0 to 7) from the latent space model. 
(d) shows $\alpha$ estimates between the Rasch model (Rasch, X-axis) and the latent space model (LatentR, Y-axis).
%The $\beta$ estimates from the Rasch model are smaller by about 2 compared with the estimates from the proposed model. 
% The rank-order correlation is 0.95. 
}
\end{figure}

We compared our parameter estimates with those from the Rasch model (Equation \ref{eq:1}). The Rasch model was estimated with the fully Bayesian approach with the same set of priors as our model's for the $\beta_i$, $\alpha_j$, $\sigma^2$ parameters. Estimation details were provided in Appendix D of the supplement.\footnote{The MCMC estimates of the Rasch model were very similar to the ML estimates obtained from the R \textsf{lme4} package \citep{bates:15}. The results were also shown in  Appendix D of the supplement.}

From the Rasch model, $\sigma$ was estimated as 2.40 (posterior median, with 95\% posterior credible interval [1.87, 3.02]), similar to the latent space IRT model estimate (2.34 with [2.07, 2.62]). 
The item parameter estimates ($\beta_i$) are displayed in Figure \ref{compare_abortion}(a) and (b). %Clearly, there is a similar pattern between the two models in terms of the item parameter estimates. 
The item parameter estimates from the Rasch model appear smaller by a constant (approximately 2 in the logit scale) compared with the latent space model, which is sensible given that the proposed model has the additional penalty term (distances). 
% whose contribution was estimated as $\gamma=2.38$. 

%> cor(rasch.m, latent.m, method="spearman")
%[1] 0.9561609
%> cor(rasch.m, latent.m, method="pearson")
%[1] 0.9978664

We then compared  the person parameter estimates ($\alpha_j$) in Figure \ref{compare_abortion}(c) and (c). 
Overall, 
the estimates under these two models are similar,
with a rank order correlation of .95. 
%Though,  respondents with the same the $\theta$ values from the Rasch model may get  different estimates from our latent space IRT model (which is understandable given the heterogeneity additionally modeled with our approach). 
\textcolor{black}{It is noteworthy that  $\alpha_j$ rarely change much compared to the Rasch model, whereas  $\beta_i$ alters due to the added latent space term. This is another evidence  that the latent space is not the ability space, at least in this example.}

%Similarities in the item and person parameter estimates between our model and the Rasch model are assuring which also serve as validating evidence for our proposed approach. 

\paragraph{Posterior predictive checking}

\begin{figure}[htbp]
\centering
%\begin{tabular}{cc}
%(a) Extended distance model & (b) Rasch model \\
\includegraphics[width=0.4\textwidth]{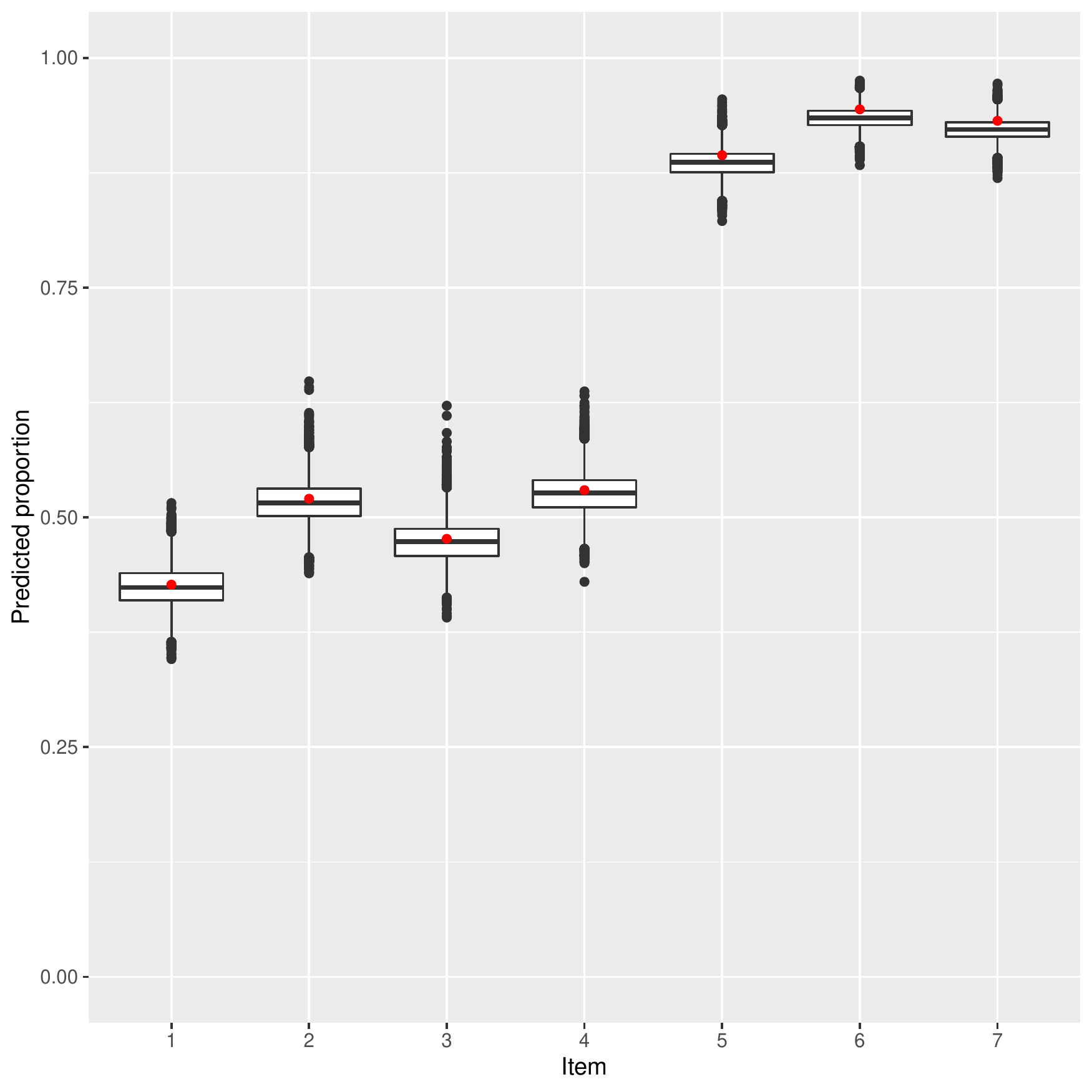} 
%\includegraphics[width=0.4\textwidth]{Posterior_checks_abortion_Rasch_new_v2.pdf} 
%\end{tabular}
\caption{\label{ppc_abortion}
Predicted proportions of the positive responses for the seven items for the attitudes to abortion data. 
%(a) is obtained from the proposed model and (b) from the Rasch model.  
%Box plots represent the distributions of the predicted proportions of positive responses in the replicated data. 
The red dot in each box indicates the proportion of positive responses calculated from the raw data. %The prediction for the latent space model is great, while the rasch model's prediction is not as good as the proposed model. 
}
\end{figure} 

We evaluated  the absolute goodness-of-fit of the proposed latent space model based on posterior predictive checking.  We compared the proportions of correct responses between the observed data and the replicated datasets (based on the estimated  model parameters). Little discrepancy between the observed and replicated measures indicates satisfactory goodness-of-fit of the model. % fit the data reasonably well.
%We compare the quality of our posterior prediction with the standard Rasch model.  
{\color{black}
Figures \ref{ppc_abortion}  displays %the results from our model (a) and from the Rasch model (b).  The box plots in each figure represent 
the box plots of the predicted proportions of positive responses for the seven items over 10,000 replicated data. The red dot in each box indicates the observed proportion. The predicted measures show highly congruent behavior to the observed measure, suggesting reasonable goodness-of-fit of the proposed model to the data under investigation. %We note in Figure \ref{ppc_abortion} (b) that the prediction is also satisfying with the Rasch model  for most items. 
Further, based on Cohen's $d$ effect size,  no item showed large mean  differences between the replicated and original data with the proposed model ($|d| > 0.8$). 
}

%while with the Rasch model, two items (Items 5 and 7) showed large mean differences compared with the original data.  

%\textcolor{red}{mj: perhaps we can put the plot back in here for the proposed model only. remove the comparison with the Rasch model (because we now have the model selection results. this comparison is not so important }

\subsection{Example 2: Deductive Reasoning}\label{sec:drv}

\subsubsection{Data and Estimation}

As a second example, 
we used the data from the Competence Profile Test of Deductive Reasoning -- Verbal assessment (DRV;  \citealp{spiel:01, spiel:08}). 
%for two reasons: 
%First, 
%the test items fall into four broad categories that are different in terms of cognitive complexity and contents.
%Therefore, 
%the dataset provides us with the opportunity to examine similarities and %Second, 
This dataset was analyzed in \citet{Jin:18},  
allowing us to compare ours to the results from the NIRM approach. 
The DRV test was developed to measure deductive reasoning of children in different developmental stages and includes 24 binary items (0 = correct, 1 = incorrect),
which fall into three broad categories:
(1) Type of inference (four levels: Modus Ponens (MP), Modus Tolens (MT), Negation of Antecedent (NA), and Affirmation of Consequence (AC)); 
(2) Content of conditional (three levels: Concrete (CO), Abstract (AB), and Counterfactual (CF)); 
and (3) Precedent of antecedent (two levels: No Negation (UN) and Negation (N)). 
More details are provided in Appendix E of the supplement.

The data include item responses from 418 school students, 
162 female and 256 male students from grade 7 to 12. 
%There was approximately the same number of students in grades 7 through 12 (age 11 through 18). 
The success rate ranged from 0.19 to 0.85 with a mean of 0.53 for the 24 test items.
The MCMC algorithm described in Section \ref{sec:estimation} was used to sample from the posterior.
\textcolor{black}{
The standard deviations of the proposal distributions were selected to ensure a reasonable acceptance rate as follows:  0.4 for $\beta$, 1.4 for $\alpha$, 0.05 for $\gamma$, 1.1 for $a$ and 0.4 for $b$. }
We generated 20,000 MCMC iterations, 
discarding the first 10,000 iterations as burn-in.
The computing time was about \textcolor{black}{56 minutes}.
The trace plots  do not show obvious signs of non-convergence (Appendix F of the supplement).
\textcolor{black}{
As an additional guard against non-convergence,
we used the Gelman-Rubin diagnostic.
To do so,
we ran the MCMC algorithm with three sets of starting values chosen at random.
We found that the scale reduction factor was less than 1.1 for all parameters,
so there were no signs of non-convergence.
The model selection method described in Section \ref{sec:model_selection} was used to determine whether the dataset was generated by the Rasch model with $\gamma = 0$ or the latent space model with $\gamma > 0$.
A posterior probability of more than $.99$ in favor of the latent space model suggests that the latent space model generated the data,
so all following results are based on the latent space model.
}

%> post.m.pi
%[1] 0.9985

% elapsed 
%2058.887

\subsubsection{Results}

\paragraph{Parameter estimates}

\begin{figure}[hptb]
\centering
\begin{tabular}{cc}
(a) $\beta_i$  & (b)  $\alpha_j$\\
\includegraphics[width=.45\textwidth]{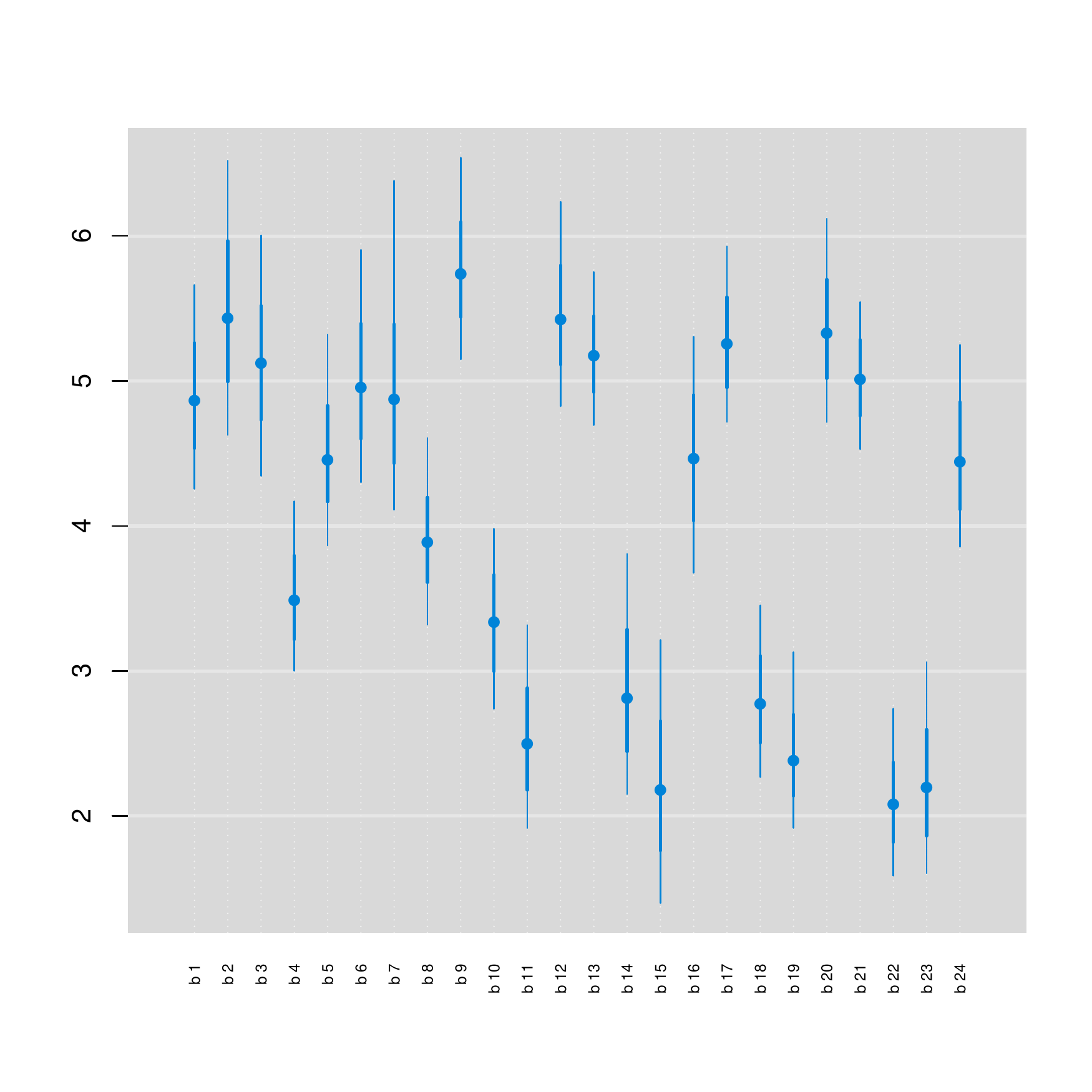} &
\includegraphics[width=.4\textwidth]{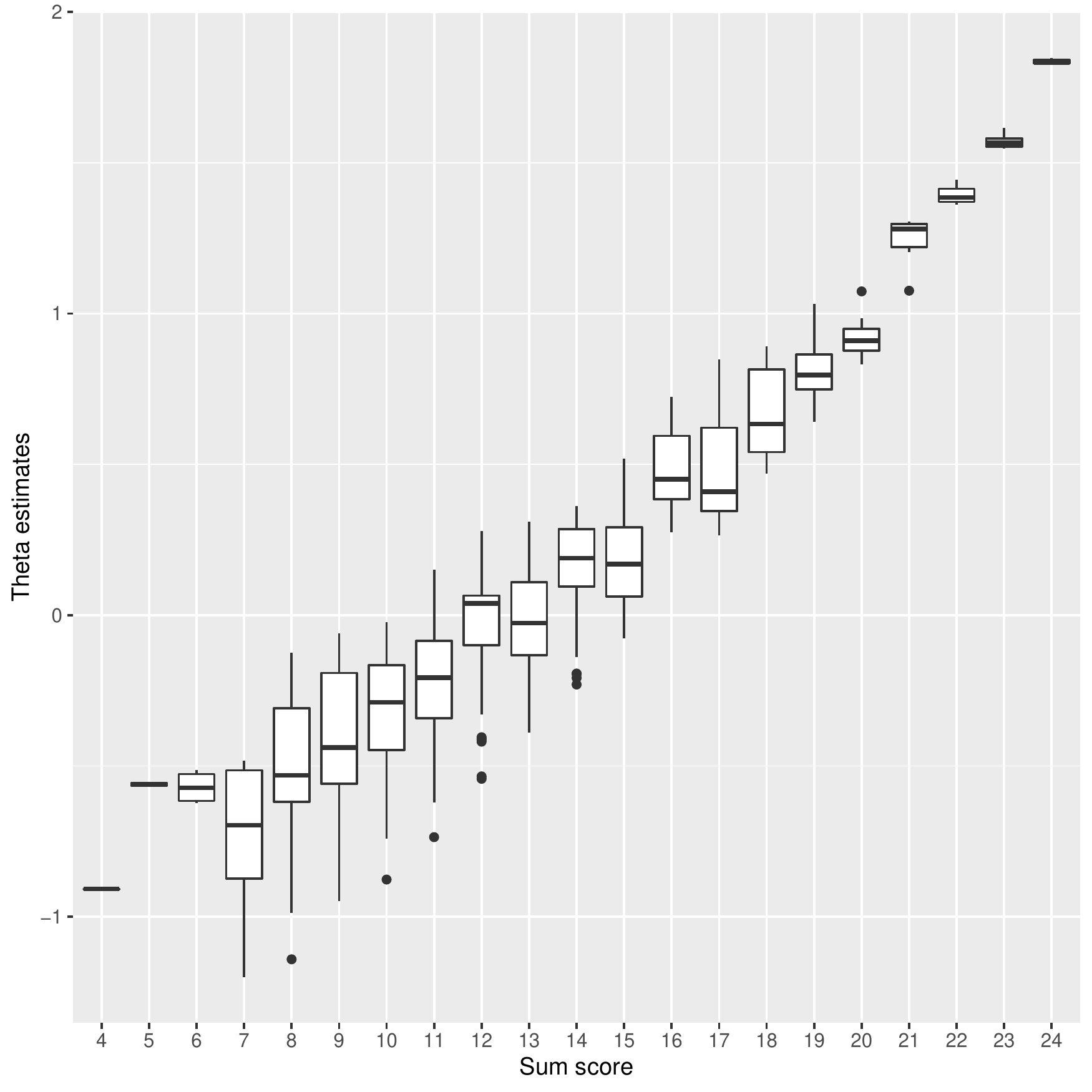} %\\
%(c) & (d) \\
%\includegraphics[width=.4\textwidth]{zw_drv_color_v2.pdf} &
%\includegraphics[width=.4\textwidth]{zw_drv_color_rotated_v2.pdf} \\ 
\end{tabular}
\caption{(a) 95\% posterior credible intervals for the $\beta_i$ estimates (b1 and b24 on the X-axis represent Items 1 to 24), and (b) the distribution of the $\alpha_j$ estimates per total test score for the DRV data. The estimates are from the latent space model. %The $\alpha_j$ estimates are nicely aligned with the sum scores. 
}\label{fg:est_drv}
\end{figure}

Figures \ref{fg:est_drv}(a) and (b) show the 95\% posterior credible intervals for the $\beta_i$ estimates of the 24 items and the distribution of the $\alpha_j$ estimates per total test score. The $\beta_i$ estimates ranged from 1 to 7 and the $\alpha_j$ estimates ranged from -2 to 2. The  $\alpha_j$ estimates were generally aliened well with the total scores. % except for a couple of lower scores (such as 4 and 5). [question: reasons?]
The latent position estimates, posterior means and 95\% posterior credible intervals, are provided in Appendix G of the supplement. 
The $\gamma$ parameter was estimated as 2.23 (posterior median, with 95\% posterior credible interval [2.08, 2.35]) and $\sigma$ as 2.51 [2.27, 2.76].  

\paragraph{Posterior predictive checking}

\begin{figure}[htbp]
\centering
%\begin{tabular}{cc}
%(a) Extended distance model &   (b) Rasch model \\
\includegraphics[width=.4\textwidth]{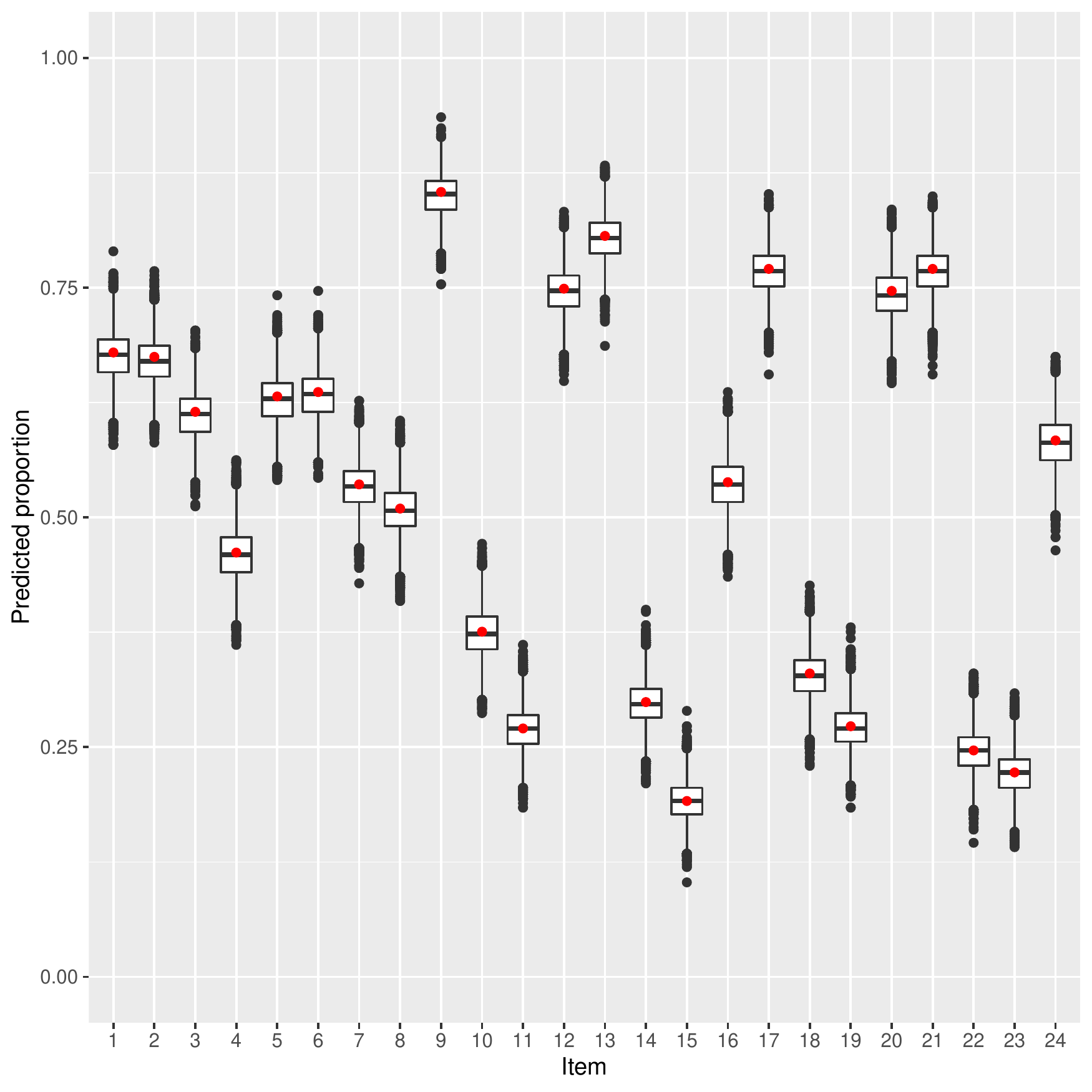}  %&
%\includegraphics[width=.4\textwidth]{posterior_checks_drv_Rasch_v2.pdf} 
%\end{tabular}
\caption{Box plots of the predicted proportions of the correct responses for the 24 DRV test items from 10,000 replicated data. The red dot in each box indicates the proportion of the correct responses for the corresponding item from the raw data.  
%(a) is from our latent space model and (b) is from the Rasch model. 
%The prediction is generally excellent with the proposed model for most items, but it is not as good  with the Rasch model for several items. 
%The Rasch model was estimated with MCMC with the same priors for $\beta_i$, $\alpha_j$, and $\sigma^2$ as the proposed model.
}\label{fg:drv_checking}
\end{figure}

{\color{black}
Goodness of fit of the latent space  model was evaluated with posterior predictive checking. % in comparison with the Rasch model which was estimated with MCMC with the same MCMC settings and the same priors for $\beta_i$, $\alpha_j$, and $\sigma^2$ as the proposed model.  
Figure  \ref{fg:drv_checking} displays the box plots of the predicted correct response proportions over 10,000 replicated responses  for the 24 DRV test items from the proposed model. The red dot in each box indicates the correct response proportion from the original data. 
The result shows that the prediction of our proposed model was excellent, supporting satisfying goodness of fit of the proposed model. %For the Rasch model, the prediction was also quite good as the proposed model for most items. 
Based on Cohen's $d$ effect size, no item showed large mean differences from the original data with the proposed model ($|d| > 0.8$). %, whereas  eleven items (Items 1, 5,  6,  7, 11, 12, 13, 16, 20, 21, 22) showed large mean differences with the Rasch model compared with the original data. 
%\textcolor{red}{mj: perhaps we can put the plot back in here for the proposed model only. remove the comparison with the Rasch model (because we now have the model selection results. this comparison is not so important }
}

\paragraph{Item structure}

%  \begin{figure}[htbp]
%  \centering
%  \begin{tabular}{cc}
% (a) & (b) \\
%  \includegraphics[width=.5\textwidth]{zw_drv.pdf} &
%  \includegraphics[width=.5\textwidth]{zw_drv1.pdf} \\
%   \end{tabular}
% \caption{Latent space for the DRV data }\label{fg:zw_drv}
%  \end{figure}

\begin{figure}[t]
\centering
\begin{tabular}{cc}
(a) DRV latent space & (b) with item group vectors \\
\includegraphics[width=.4\textwidth]{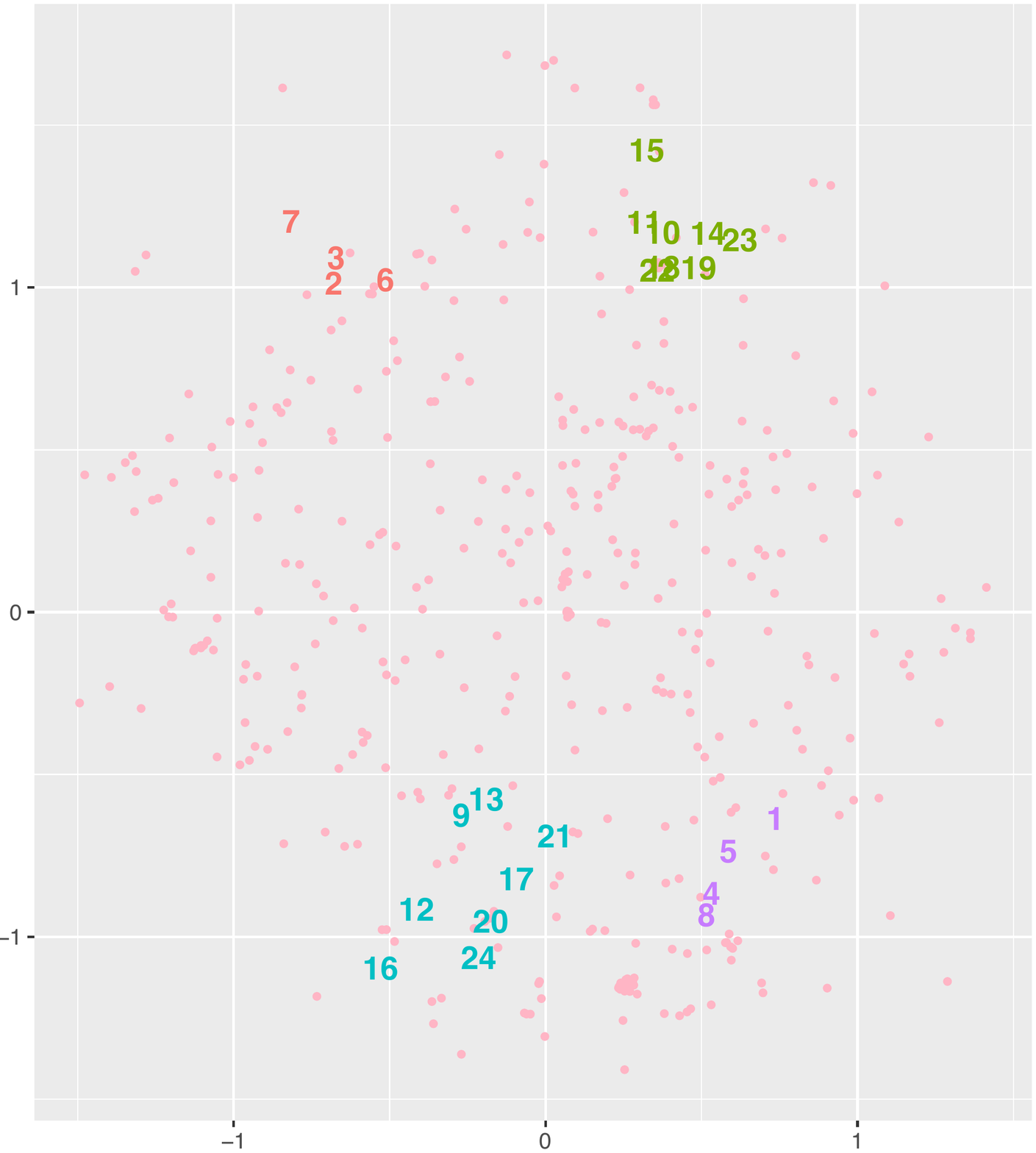} &
\includegraphics[width=.4\textwidth]{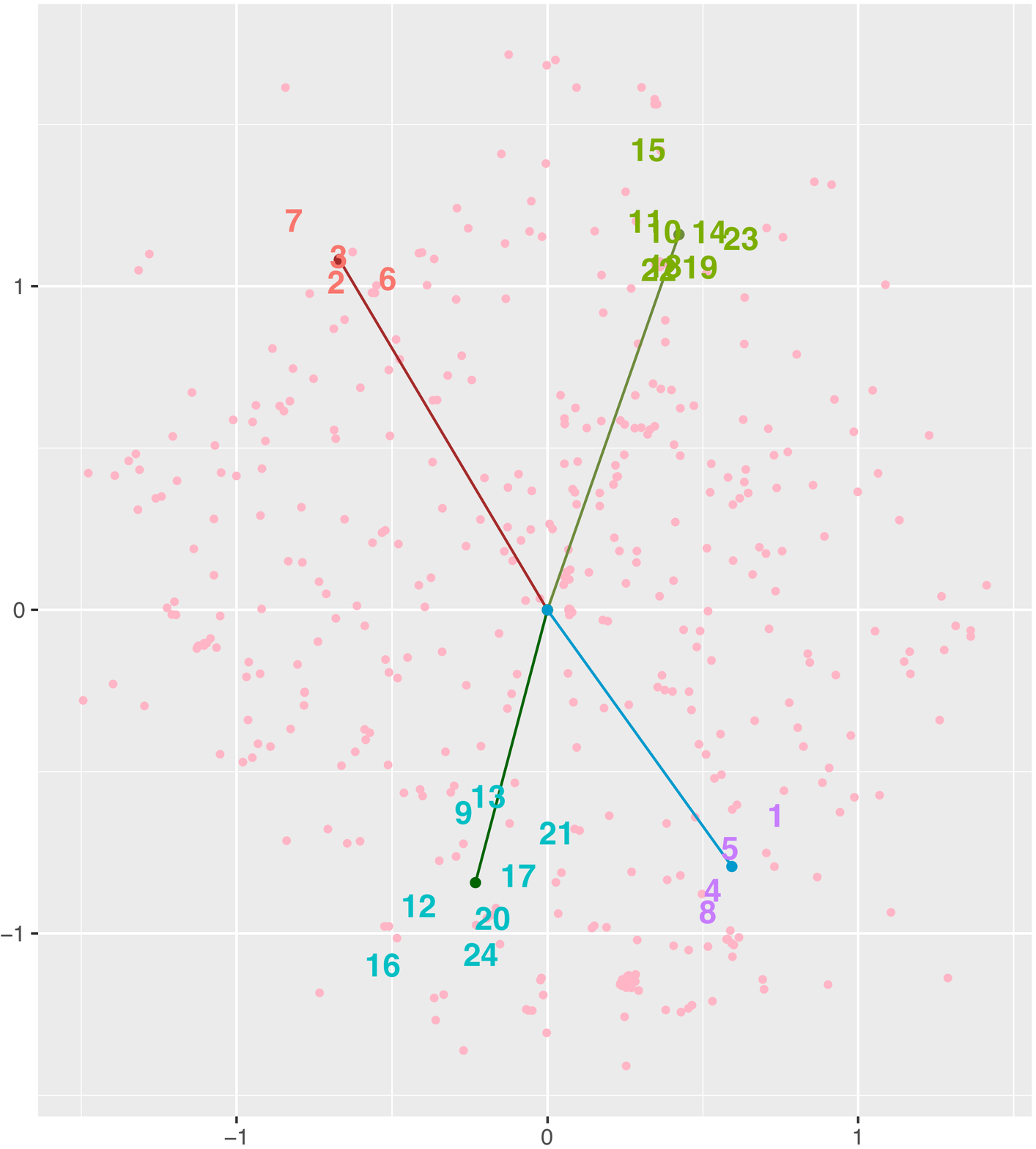} \\ 

\end{tabular}
\caption{(a) Latent space for the DRV data and (b) DRV data latent space superimposed with the four vectors that represent the centers of four item groups (I1, I2, I3, and I4). 
%and rotated latent space with oblim rotation (b). 
In both figures, dots represent respondents and numbers represent items. Four item groups are distinguished with four different colors. 
I1: Items 2,3,6,7;
I2: Items 10,11,14,15,18,19,22,23;
I3: Items 9,12,13,16,17,20,21,24;
I4: Items 1,4,5,8. 
}\label{fg:zw_drv}
\end{figure}

Figure  \ref{fg:zw_drv}(a) displays the estimated latent space, where bullet points  represent respondents and numbers represent items. Roughly, four item groups appear as  color-coded for distinction. The four item group members are listed in Table \ref{tab:drv_items}. The item structure identified here 
shows an excellent agreement with the structure identified in \citet{Jin:18} based on the NIRM approach. %This congruence serves as convergent validity evidence. 

\begin{table}[t]
\centering
\begin{tabular}{ c|c } 
 \hline
 Item group & Group details  \\ 
  \hline 
 I1 & %2,3,6,7
UN\_CO\_NA (2);
UN\_CO\_AC (3);
N\_CO\_NA (6);
N\_CO\_AC (7)

  \\ 
\hline 
I2 & %10,11,14,15,18,19,22,23
UN\_AB\_NA (10);
UN\_AB\_AC (11);
N\_AB\_NC (14);
N\_AB\_AC (15); \\
& UN\_CF\_NA (18);
UN\_CF\_AC (19);
N\_CF\_NA (22);
N\_CF\_MT (23)
\\

 \hline 
I3 & %9,12,13,16,17,20,21,24
UN\_AB\_MP (9);
UN\_AB\_MT (12);
N\_AB\_MP (13);
N\_AB\_MT (16); \\ 
& UN\_CF\_MP (17);
UN\_CF\_MT (20);
N\_CF\_MP (21);
N\_CF\_MT (24)
\\
 \hline 

 I4 & %1,4,5,8
UN\_CO\_MP (1);
UN\_CO\_MT (4);
N\_CO\_MP (5);
N\_CO\_MT (8) \\
 \hline
 \end{tabular}
\caption{Members of the four item groups identified in the DRV data latent space. Numbers in parenthesis indicate item numbers. The acronyms in the item labels indicate the following design factors and their levels: (1) UN vs. N:  no negation (UN) and Negation (N) for the presentation of the antecedent factor. (2) CO vs. AB vs. AC: Concrete (CO), Abstract (AB), and Counterfactual (CF) for the content of conditional factor.  (3) MP vs. MT vs. NA vs. AC: Modus Ponens (MP),  Modus Tollens (MT), Negation of Antecednet (NA), and Affirmation of Consequent (AC) for the type of inference factor. 
}
\label{tab:drv_items}
\end{table}

I1 and I2 in the upper part  of the latent space consist of Concrete items (CO). They are further differentiated in terms of Type of Inference; I1 on the left includes bi-conditional inference items (MP and MT) and I2 on the right includes more complex inference type items (NA and AC). I3 and I4 in the bottom part of the latent space consist of logical fallacy items (Ab and CF).   They are further separated by Type of Inference; I3 on the left includes bi-conditional items (MP and MT) and I4 on the right includes complex algebra items (NA and AC). The Presentation of Antecedent factor (UN vs. N) is mixed in all groups, meaning that this factor hardly contributes to item differentiation. 

\paragraph{Success probabilities for item group}

\begin{figure}[h]
\centering
\begin{tabular}{cc}
(a) Item group l1 & (b) Item group l2\\
\includegraphics[width=.3\textwidth]{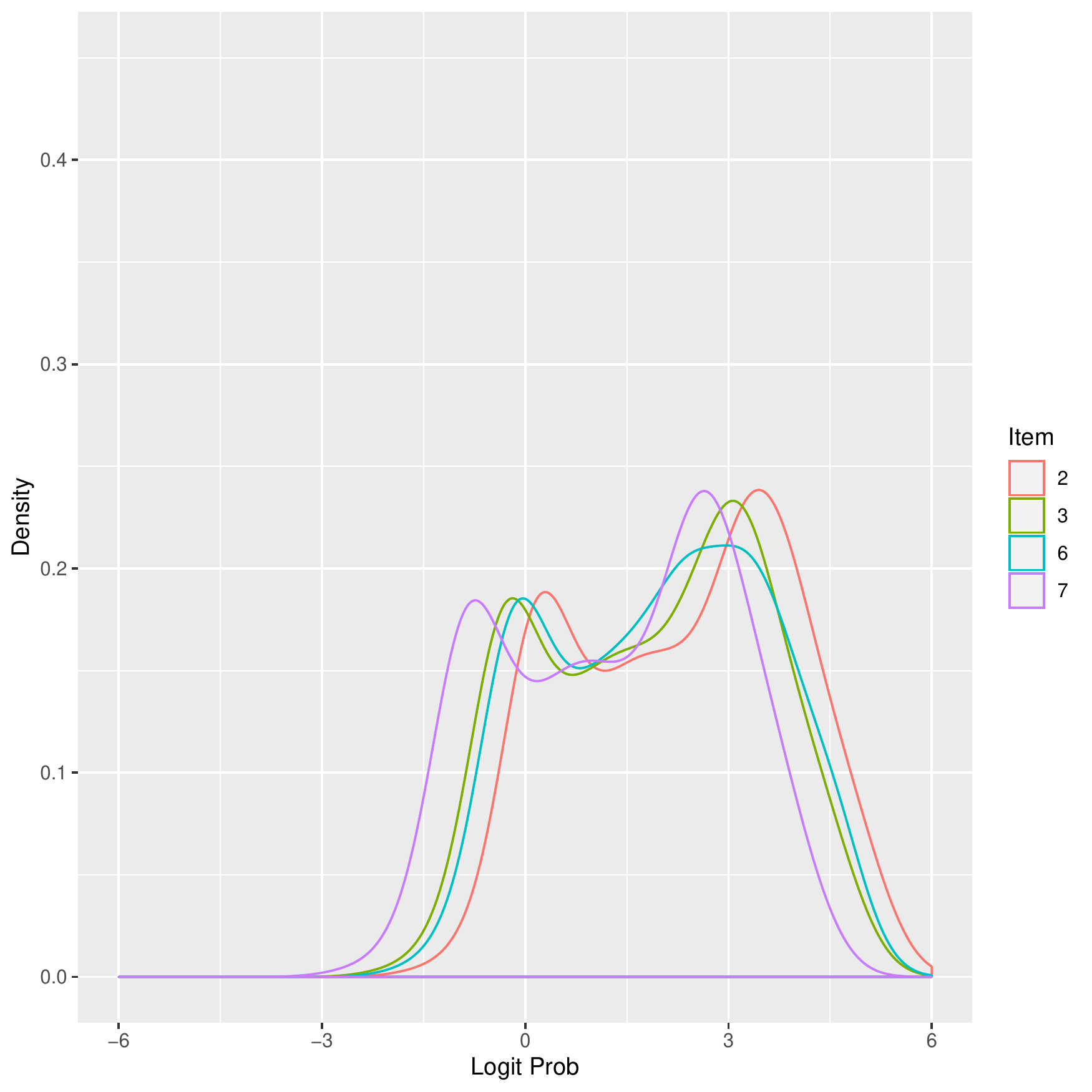} &
\includegraphics[width=.3\textwidth]{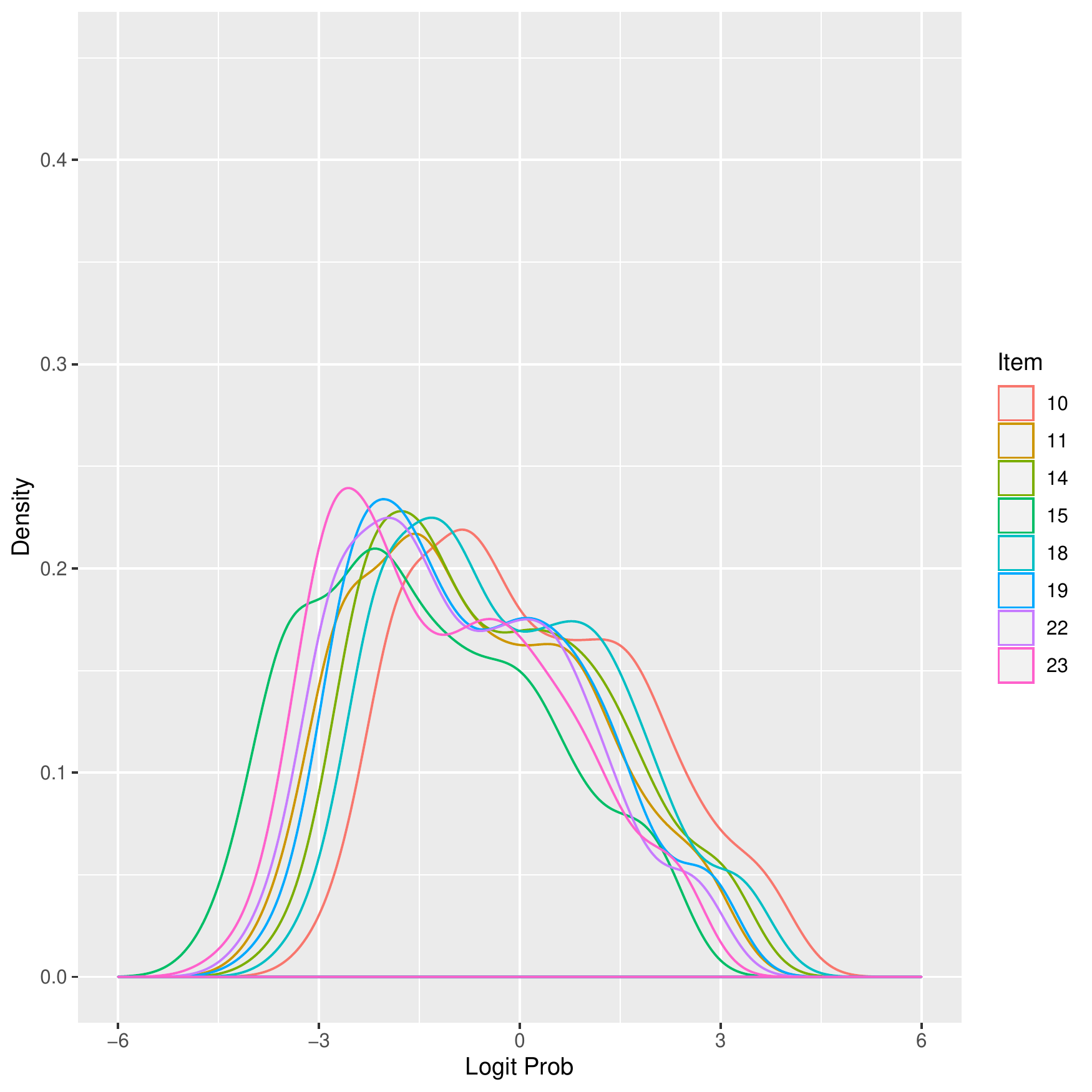} \\
(c) Item group l3 & (d) Item group l4 \\
\includegraphics[width=.3\textwidth]{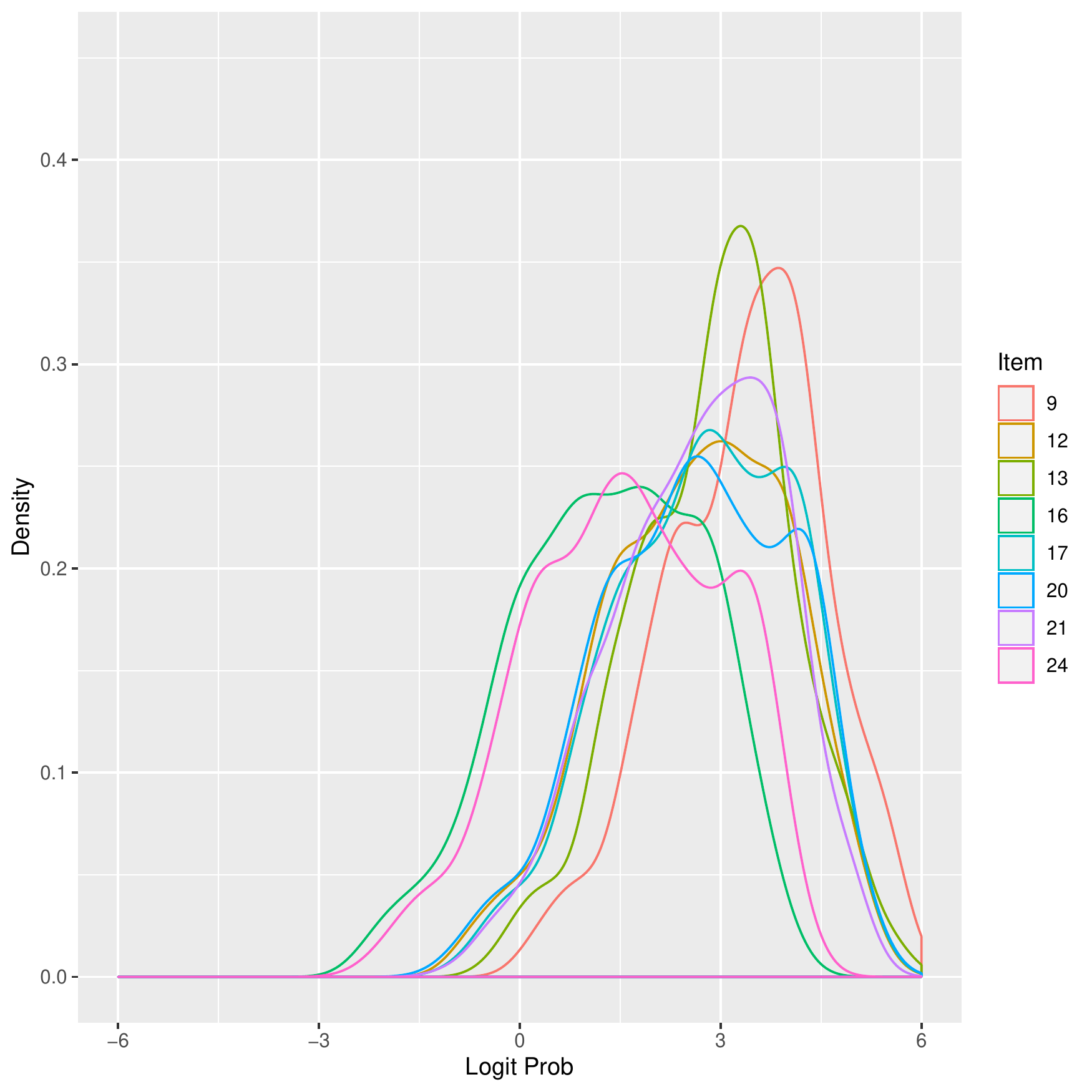} &
\includegraphics[width=.3\textwidth]{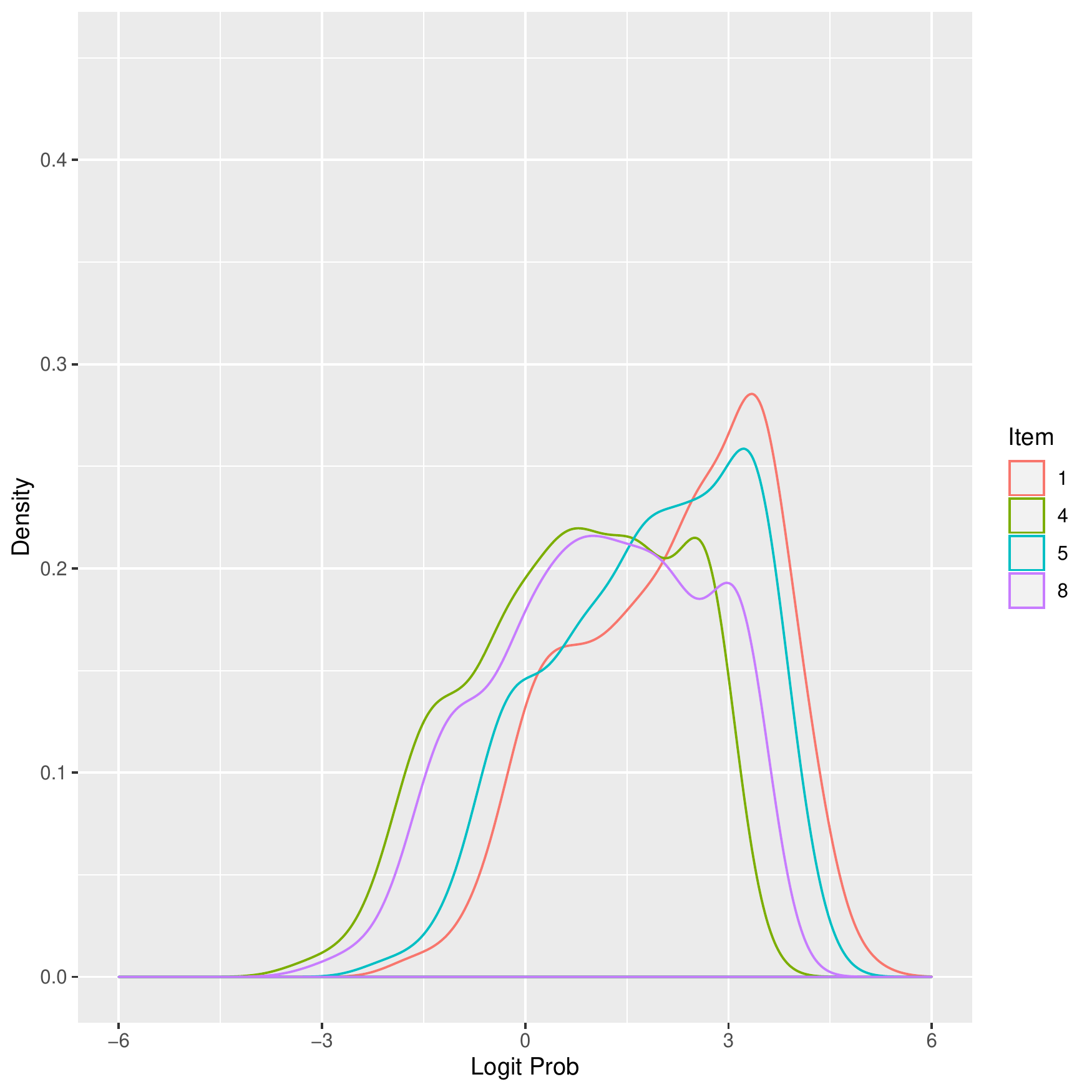} \\ 
\end{tabular}
\caption{Density plots of the log odds success probabilities of the individual DRV test items per item group. (a) to (d) represent the density plots for item groups I1 to I4 groups in order. %The logit success probabilities are similar within item group but different across item groups. I1: Items 2,3,6,7;
I2: Items 10,11,14,15,18,19,22,23;
I3: Items 9,12,13,16,17,20,21,24;
I4: Items 1,4,5,8. 
}\label{fg:drv_items}
\end{figure}

We assessed the correct response probabilities within and between the four identified item groups (I1, I2, I3, and I4). The density plots of the log odds success probabilities of the individual items per item group are presented in Figure \ref{fg:drv_items}. 
While the four groups show different patterns of the logit success probabilities, items within the same group are similar in the patterns. %I3 and I4 show  mixed patterns, meaning that they have more heterogeneous in characteristics than the I1 and I2 item members. % (I3 and I4 items are less close to each other than I1 and I2 items). 

%  post.m.b[c(1,4,5,8)]
%       V1       V4       V5       V8 
% 4.046867 2.883745 3.718051 3.170981 

% > post.m.b.low[c(1,4,5,8)]
%       V1       V4       V5       V8 
% 3.514197 2.412898 3.244810 2.679165 
% > post.m.b.up[c(1,4,5,8)]
%       V1       V4       V5       V8 
% 4.697918 3.474466 4.334734 3.725752 

\paragraph{Cosine similarity between item groups}

We evaluated similarities between two positions in a latent space by the cosine similarity measure. 
%One  idea is to use the cosine of the angle between two non-null vectors that represent positions of items or respondents in the estimated latent space. 
The cosine similarity of two vectors $\bm{a} \in \mR^p$ and $\bm{b} \in \mR^p$ of length $||\bm{a}||_2 > 0$ and $||\bm{b}||_2 > 0$ was  computed as 
\[
\begin{array}{ccc}
\cos(\theta) 
&=& \dfrac{\bm{a}^\top\, \bm{b}} {||\bm{a}||_2\;  ||\bm{b}||_2},
\end{array}
\] 
where $\theta$ is the angle between two vectors $\bm{a}$ and $\bm{b}$.
The cosine similarity measure takes on values in the interval $[-1,1]$. Two vectors pointing into the same direction have a cosine similarity of 1,
two vectors with opposite directions have a cosine similarity of -1,
and two orthogonal vectors have a cosine similarity of 0.

While cosine similarity can be computed between any two positions in a latent space -- including positions of items, respondents, and both items and respondents -- we focus here on similarities between the four item groups. 
Figure  \ref{fg:zw_drv}(b) shows the original DRV latent space added with the four vectors indicating the centers of the four item groups (where the centers are the mean positions of the corresponding item group members).  
Table \ref{tab:similarity} presents a matrix of cosine similarity measures between the four item groups. 

\begin{table}[t]
\centering
\begin{tabular}{ c c c c c } 
 \hline
   & I1    & I2 & I3 & I4  \\
\hline 
I1 &   -    &      & & \\
I2 & 0.618 &   -   & & \\
I3 & -0.680&-0.996& -& \\
I4 & -0.996& -0.546&0.613 & -\\
\hline 
 \end{tabular}
\caption{Cosine similarity measures between (centers of) the four item groups. I1: Items 2,3,6,7;
I2: Items 10,11,14,15,18,19,22,23;
I3: Items 9,12,13,16,17,20,21,24;
I4: Items 1,4,5,8.}
\label{tab:similarity}
\end{table}

% > cos_sim12
%       cos 
% 0.6185891 
% > cos_sim13
%       cos 
% -0.6806504 
% > cos_sim14
%       cos 
% -0.996085 
% > cos_sim23
%       cos 
% -0.996664 
% > cos_sim24
%       cos 
% -0.5467093 
% > cos_sim34
%       cos 
% 0.6132223 
% > 

%  Item group 1  & Item group 2 &  Cosine similarity  \\ 
%   \hline 
%  1 & 2 &  0.590 \\
%   & 3 &  -0.687 \\
%   & 4 &  -0.995 \\
%  \hline 
%  2 & 3 & -0.992  \\
%   & 4 & -0.508  \\ 
%  \hline 
%  3 & 4 &  0.613 \\ 
%  \hline

Table \ref{tab:similarity} confirms that I1 is most dissimilar to I4 and I2 is most dissimilar to I3. 
%Similarity between Item groups 1 and 2 is alike to the similarity between Item groups 3 and 4. 
Marked dissimilarities  between I1 and I4 and between I2 and I3 support our earlier finding that  Type of Inference (MP/MT vs. NA/AC)  most substantially differentiates the DRV test items. 

% cosign similarity 
% > cos_sim12
%       cos 
% 0.5901487 
% > cos_sim13
%       cos 
% -0.6871853 
% > cos_sim14
%       cos 
% -0.9952579 
% > cos_sim23
%       cos 
% -0.9920266 
% > cos_sim24
%       cos 
% -0.5088237 
% > cos_sim34
%       cos 
% 0.6132609 

\paragraph{Respondent structure}

%We evaluated respondents' performance by their relatives distances to individual test items. 
%To illustrate, 
To evaluate their performance in the latent space, we first categorized the children into four sub-groups based on their proximity to the four item groups: % that we identified: 
\begin{enumerate}
\item [(1)] Children near I1. They performed well on logical fallacy inference items (NA/AC) but poorly with simpler inference items (MP/NT) when the items involved  
concrete conditionals (Co); 
\item [(2)] Children near I2. They performed well on logical fallacy inference items (NA/AC) but poorly with simpler inference items (MP/NT) if the items involved abstract or counterfactual conditionals  (NA/AC);  
\item [(3)] Children near I3. They performed well on simpler inference items (MP/MT) but poorly with logical fallacy inference items (NA/AC) if the items involved abstract or counterfactual conditionals  (NA/AC); 
\item [(4)] Children near I4. They performed well on simpler inference items (MP/MT) but poorly with logical fallacy inference items (NA/AC) if the items involved concrete conditionals (Co); 
\end{enumerate}

%Children in sub-groups 1 and 2 are close to each other but they are apart from children in sub-groups 3 and 4. 
%Based on the two groups' performance, it may be reasonable to conjecture that 
Based on the above, we can reasonably  conclude that children in sub-groups 3 and 4 were at a lower level of deductive reasoning than those children in sub-groups 1 and 2. %e latter children were able to handle more complex inference items than the former).
While performing well with complex inference items, children in sub-groups 1 and 2 showed poor performance on simpler inference items. This indicates that they might be in a transition to a higher developmental stage. Children in a transition stage tend to make mistakes with easier items, for instance, due to over-generalization on simple problems \citep[e.g.,][]{markovits:98, draney:07} %, Fleury, Quinn, & Venet, 1998; Draney, 2007).

Further, it is possible to make additional sub-grouping of children. For instance, % as follows: %Additional sub-grouping of children is possible. For instance, 
\begin{enumerate}
% \item [(5)] 
% Children located between I3 and I4 were good with simple inference items (MP/MT) but not with logical fallacy items (NA/AC) with either concrete or complex (abstract/counterfactual) conditionals; 
% \item [(6)]  Children located between I1 and I2  were good with logical fallacy items (NA/AC)  but not with simple inference items (MP/MT) that involve either concrete or complex (abstract/counterfactual) conditionals. 
\item [(5)] Children between I1 and I3. They were good with both simple and logical fallacy inference items when the items were combined with abstract/counterfactual and concrete conditionals, respectively; 
\item [(6)] Children between I2 and I4. They were good with  both simple and logical fallacy inference items when the items were combined with concrete and abstract/counterfactual conditionals, respectively; 
\item [(7)]  Children around the center of the latent space. They performed equally well on most test items. %\textcolor{black}{[mj: should we mention that there can be losers as well as winners in  this area? luckily no  losers in this dataset]}
\end{enumerate}

How can we identify a specific sub-group for each respondent? One could draw a contour that represents a 95\% posterior credible region for each child.
If the contours of children overlap, 
the children may form a subgroup of children that are similar.  In addition, 
one can directly calculate a distance between an individual respondent and each item or item group. For instance, suppose respondent A has a distance of 0.5 from I1, 1.5 from I2, 2 from I3, and 3 from I4. %This means that respondent A is closest to I1, compared with other item groups 
That is, the ratio or relative distances is 1:3:4:, meaning that respondent A belongs to the subgroup near I1. So, it is possible to categorize respondents based on their quantified relative distances.

\paragraph{Comparisons with principal component analysis and factor analysis}

Our model showed that items in the same item group are similar, while items in different  groups are distinctive in terms of contents as well as  success probabilities. 
Traditional methods, such as principal component analysis (PCA) and factor analysis (FA), might be used for  
similar purposes. 
%While PCA and FA have differences in underlying models and estimation principles, but they both purpose to identify groups of items that are similar with respect to the principal components or factors that are extracted from the data being analyzed. 

\begin{figure}[t]
\centering
\begin{tabular}{cc}
(a) PCA  & (b) Factor analysis \\
\includegraphics[width=.4\textwidth]{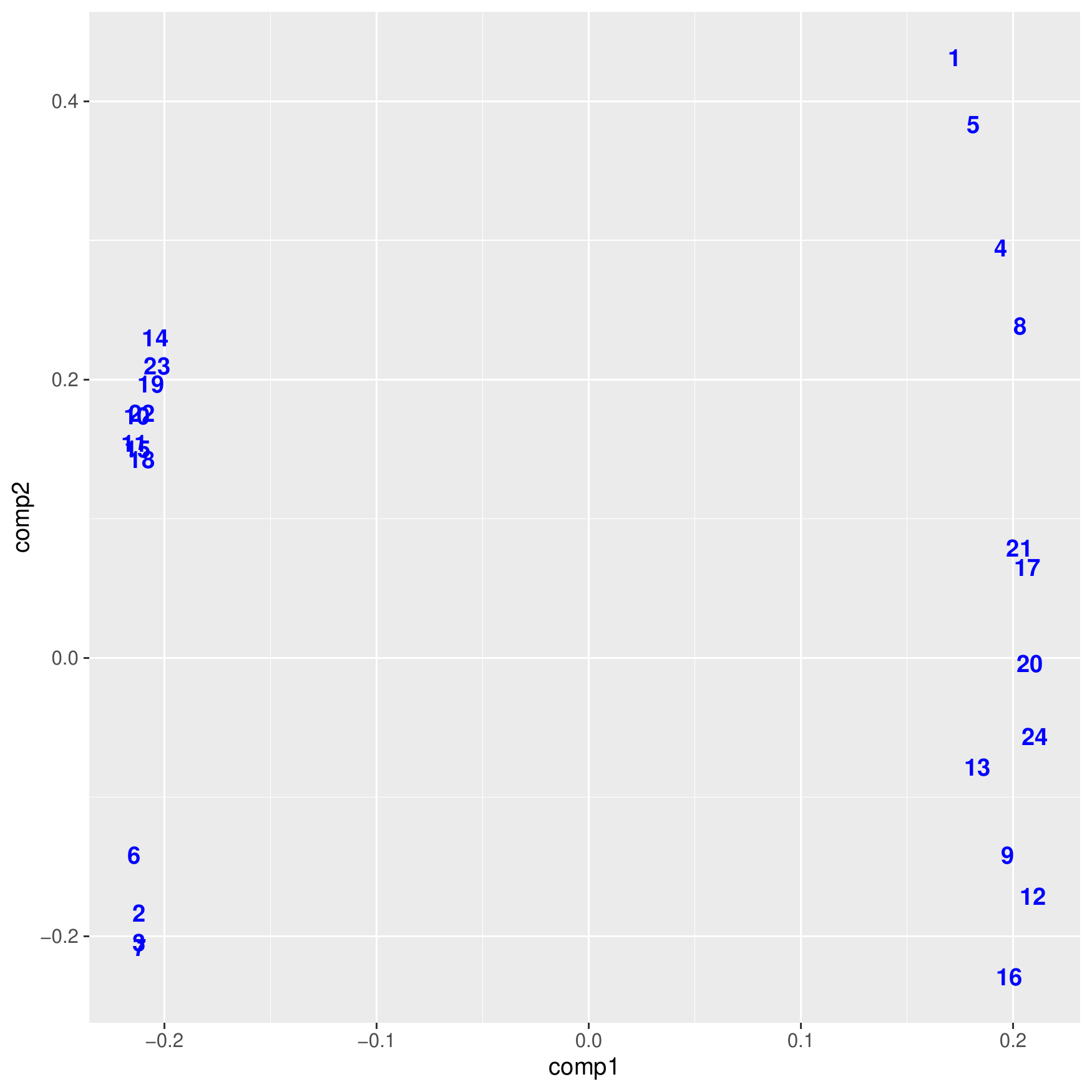} &
\includegraphics[width=.4\textwidth]{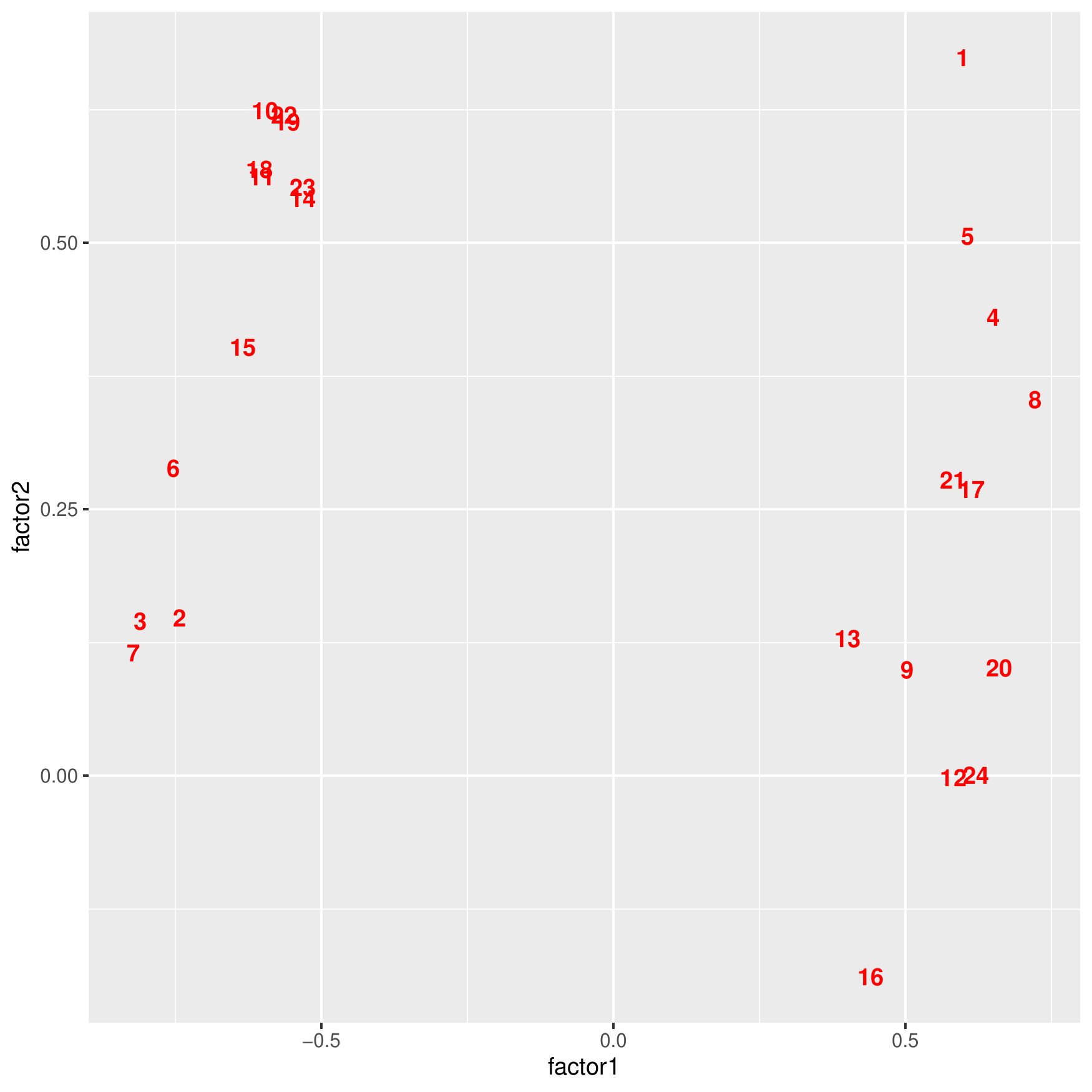} 
\end{tabular}
\caption{(a) Principal component analysis (PCA) solution with a tetrachoric correlation matrix as input and (b) item factor analysis (FA) solution with oblim rotation for the DRV data. %Item grouping (four sub-groups) from PCA and FA are roughly similar to the grouping identified with our proposed approach. The grouping is less clear with the FA approach.  
}\label{fg:pca}
\end{figure}

To compare, we applied PCA and FA to the DRV data where two principal components and two factors were extracted.\footnote{For PCA, a tetrachoric correlation matrix was used as input data with the R \textsf{psych} package \citep{revelle:19}. For FA, item factor analysis is applied with oblim rotation by using the R mirt \textsf{package} \citep{chalmers:12}. With both methods, two-dimensional solutions were  optimal. }
The solutions from the two methods are presented in Figure \ref{fg:pca}. Items are placed in the two-dimensional spaces that represent the two principal components or  factors. 
The item clusters identified with PCA and FA are roughly similar to our approach. With factor analysis, the membership of a few items, such as Items 6, 15, 17, and 21, were less clear compared with the other approaches.  
%Similarity with the solutions from these traditional methods is assuring. % serves as additional convergent validity evidence for our approach.  
%Though, it would be useful to point out  some important differences between the methods. % between our approach and the  methods. % differences.  %between our approach and PCA/FA. 
However, two important differences need to be clarified: (1) in the two traditional methods, extracting factors or principal components (dimensions) is often the main interest, whereas it is not the case in our approach; (2) both in PCA and FA, items and respondents cannot be placed in the same space, unlike our approach.

% \begin{table}[ht]
% \centering
% \begin{tabular}{rrr|rrr}
%   \hline
%   & \multicolumn{2}{c|}{Two factor} & \multicolumn{3}{c}{Three factor } \\   
%  & Factor1 & Factor2 & Factor1 & Factor2 & Factor3 \\ 
%   \hline
%   V1 & 0.60 & 0.67 & 0.43 & 0.68 & -0.41 \\ 
%   V2 & -0.74 & 0.15 & -0.35 & 0.14 & 0.75 \\ 
%   V3 & -0.81 & 0.14 & -0.38 & 0.12 & 0.84 \\ 
%   V4 & 0.65 & 0.43 & 0.50 & 0.45 & -0.40 \\ 
%   V5 & 0.61 & 0.51 & 0.43 & 0.52 & -0.41 \\ 
%   V6 & -0.75 & 0.29 & -0.30 & 0.30 & 0.83 \\ 
%   V7 & -0.82 & 0.12 & -0.41 & 0.09 & 0.83 \\ 
%   V8 & 0.72 & 0.35 & 0.56 & 0.37 & -0.43 \\ 
%   V9 & 0.50 & 0.10 & 0.51 & 0.12 & -0.15 \\ 
%   V10 & -0.60 & 0.62 & -0.44 & 0.61 & 0.42 \\ 
%   V11 & -0.60 & 0.56 & -0.45 & 0.54 & 0.44 \\ 
%   V12 & 0.58 & -0.00 & 0.62 & 0.01 & -0.14 \\ 
%   V13 & 0.40 & 0.13 & 0.36 & 0.14 & -0.20 \\ 
%   V14 & -0.53 & 0.54 & -0.45 & 0.53 & 0.30 \\ 
%   V15 & -0.63 & 0.40 & -0.52 & 0.38 & 0.40 \\ 
%   V16 & 0.44 & -0.19 & 0.43 & -0.17 & -0.18 \\ 
%   V17 & 0.61 & 0.27 & 0.78 & 0.32 & 0.04 \\ 
%   V18 & -0.61 & 0.57 & -0.64 & 0.58 & 0.15 \\ 
%   V19 & -0.56 & 0.61 & -0.62 & 0.63 & 0.09 \\ 
%   V20 & 0.66 & 0.10 & 0.83 & 0.12 & 0.03 \\ 
%   V21 & 0.58 & 0.28 & 0.73 & 0.33 & 0.03 \\ 
%   V22 & -0.56 & 0.62 & -0.54 & 0.63 & 0.21 \\ 
%   V23 & -0.53 & 0.55 & -0.63 & 0.58 & 0.00 \\ 
%   V24 & 0.62 & 0.00 & 0.78 & 0.00 & 0.00 \\ 
%   \hline
% \end{tabular}
% \caption{Factor analysis solutions with two and three factors. Oblim rotation was applied.} 
% \end{table}

\paragraph{Latent space rotation}

%We discussed in Section xx that one can draw a substantively meaningful interpretation for the latent space dimensions (x-axis and y-axis) based on neighboring items. 

\begin{figure}[t]
\centering
\includegraphics[width=.4\textwidth]{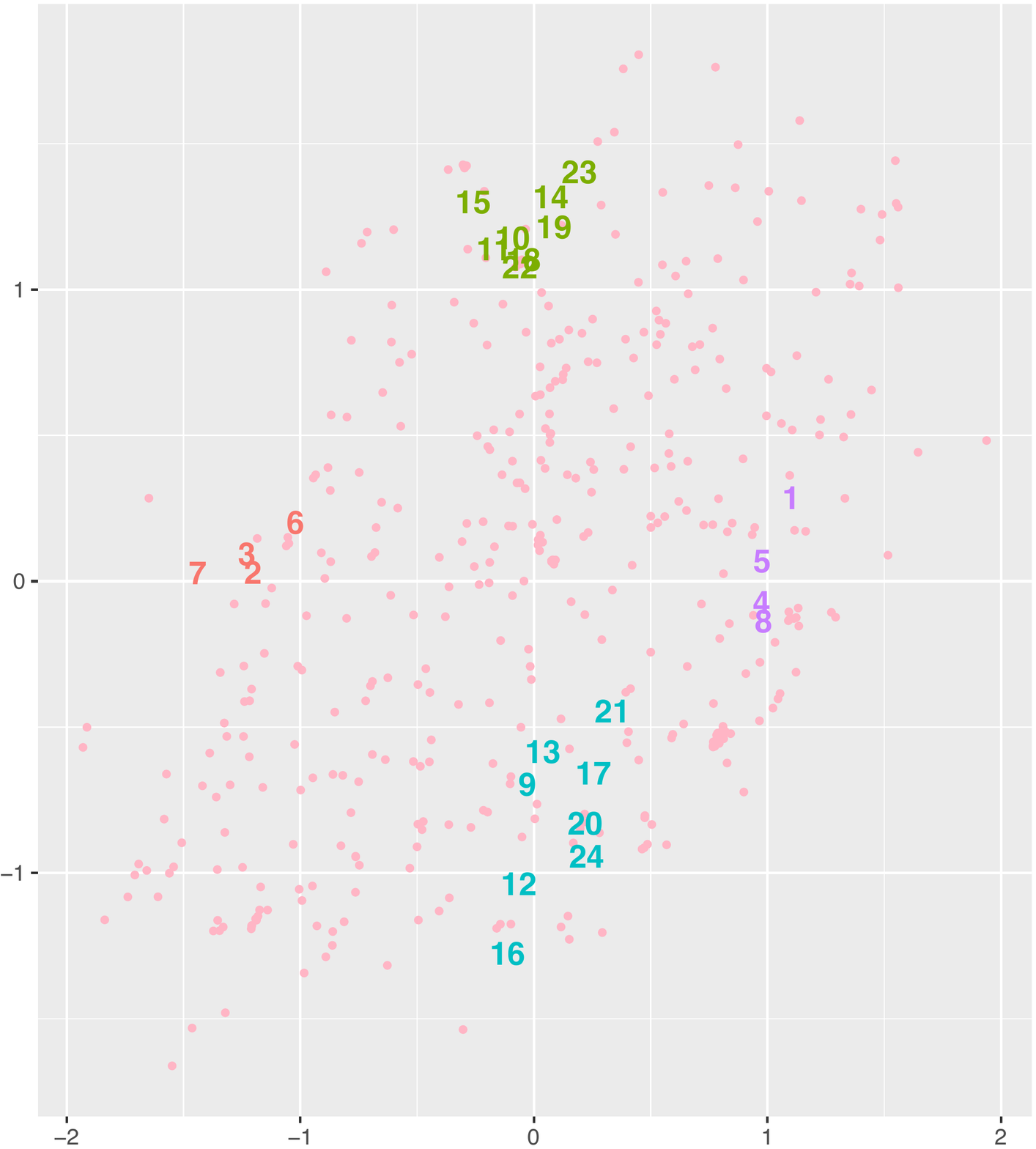}  
\caption{Rotated latent space for the DRV data with oblim rotation. Dots represent respondents and numbers represent items. Four item groups are distinguished with four different colors. %Two item groups (I1, I4) are on the X-axis and two item groups (I2, I3) are on the Y-axis. 
I1: Items 2,3,6,7;
I2: Items 10,11,14,15,18,19,22,23;
I3: Items 9,12,13,16,17,20,21,24;
I4: Items 1,4,5,8. }\label{fg:zw_drv2}
\end{figure}

%Understanding latent space dimensions is not of our main interest; however, 
When desirable, we can attach substantive interpretations to latent space dimensions based on neighboring items. To illustrate, we return to the original latent space displayed in Figure \ref{fg:zw_drv}(a). 
No items appear close to the X-axis; therefore, it is difficult to interpret the X-axis in a meaningful way. 

To improve interpretability, 
we rotate the original  latent space to a place where items are better encompassed by the axes, which is permitted due to rotational invariance property of latent space. Rotation is a frequently utilized technique in factor analysis which also has rotational invariance. %,  to improve factor interpretability.   
We applied oblim rotation \citep{Jennrich:02} to the estimated item position matrix $\bm{B}$ using the R package \textsf{GPArotation} \citep{bernaards:05},
%Then the rotated matrix  $\bm{W}^*$ is obtained as 
%\begin{equation}\label{eq:rotation}
%\bm{W}^* = \bm{W} \cdot \{\bm{T}^{-1}\}'
%\end{equation}
%where $\bm{T}$ is a rotation matrix. We 
and then rotated the respondent position matrix $\bm{B}$ in the same way with a common rotation matrix. 
%by using the rotation matrix $\bm{T}$ from Equation \eqref{eq:rotation}, and
We denote the rotated item and respondent position matrices by $\bm{A}^*$ and $\bm{B}^*$,
respectively.
%so that the respondents could be rotated in the same way as the items ($Z^*$ indicates the rotated $Z$ matrix). 

Figure \ref{fg:zw_drv2} displays 
the rotated latent space for the DRV data. %, where items and respondents are placed based on $\bm{B}^*$ and $\bm{A}^*$. 
%Observe that in this rotated space, 
Two item groups I1 and I4 are positioned close to the X-axis, while I2 and I3 are placed close to the Y-axis in the rotated space. 
This indicates that the X-axis represents Type of Inference (MP/MT vs. NA/AC) combined with Concrete conditionals, while the Y-axis represents Type of Inference (MP/MT vs. NA/AC) combined with Abstract and Counterfactual conditionals. Items are differentiated  based on the type of inference in each dimension, while the two dimensions are separated by the content of conditionals (concrete vs. abstract/counterfactual). 
%[mj: additional interpretations or implications?]

% \paragraph{Comparison between 2d and 3d}

% \begin{figure}[htbp]
% \centering
% \begin{tabular}{c}
% (a) \\ 
% \includegraphics[width=.6\textwidth]{drv_factor_analysis_3d.png} \\
% (b) \\
% \includegraphics[width=.8\textwidth]{drv_3d_v1.png} 
% \end{tabular}
% \caption{Three dimensional solutions: factor analysis (a) and latent space model (b) }\label{fg:3d}
% \end{figure}

{\color{black}
\section{Simulation Study}\label{sec:simulation}

We conducted a simulation study to evaluate whether the model selection approach described in Section \ref{sec:model_selection} can determine if the Rasch model with $\gamma=0$ or the latent space model with $\gamma > 0$ generated a given dataset. 
To do so, 
we used the setting of Figure 1(a) and 1(b) with $N=200$ and $I=14$. 
We simulated 100 datasets under the Rasch model with $\gamma = 0$ and under the latent space model with  $\gamma = 1.7$. 
For each simulated dataset, 
the MCMC algorithm was run for 5,000 burn-in iterations,
followed by 5,000 post-burn-in iterations.
We estimated the posterior probability of the indicator $\delta = 1$ by the proportion of times $\delta=1$ in a Markov chain Monte Carlo sample from the posterior. 
% We choose an uninformative prior $B (a_\omega, b_\omega)$ for $\omega$ with $a_\omega=1$ and $b_\omega = 1$. For the spike prior, the mean and standard deviation are chosen such that the density is concentrated near zero ($\mu_{\gamma_0} = -3$, $\tau_{\gamma_0}=1$). For the slab prior, the general prior is used that we selected for $\gamma$ in this paper ($\mu_{\gamma_0} = 0.5$, $\tau_{\gamma_0}=1$). %\textcolor{red}{mj: should we show the prior distribution? I have a graph}

\begin{figure}[phtb]
\centering
\begin{tabular}{cc}
(a) Truth: $\gamma = 0$  & (b) Truth: $\gamma = 1.7$\\
\includegraphics[width=.4\textwidth]{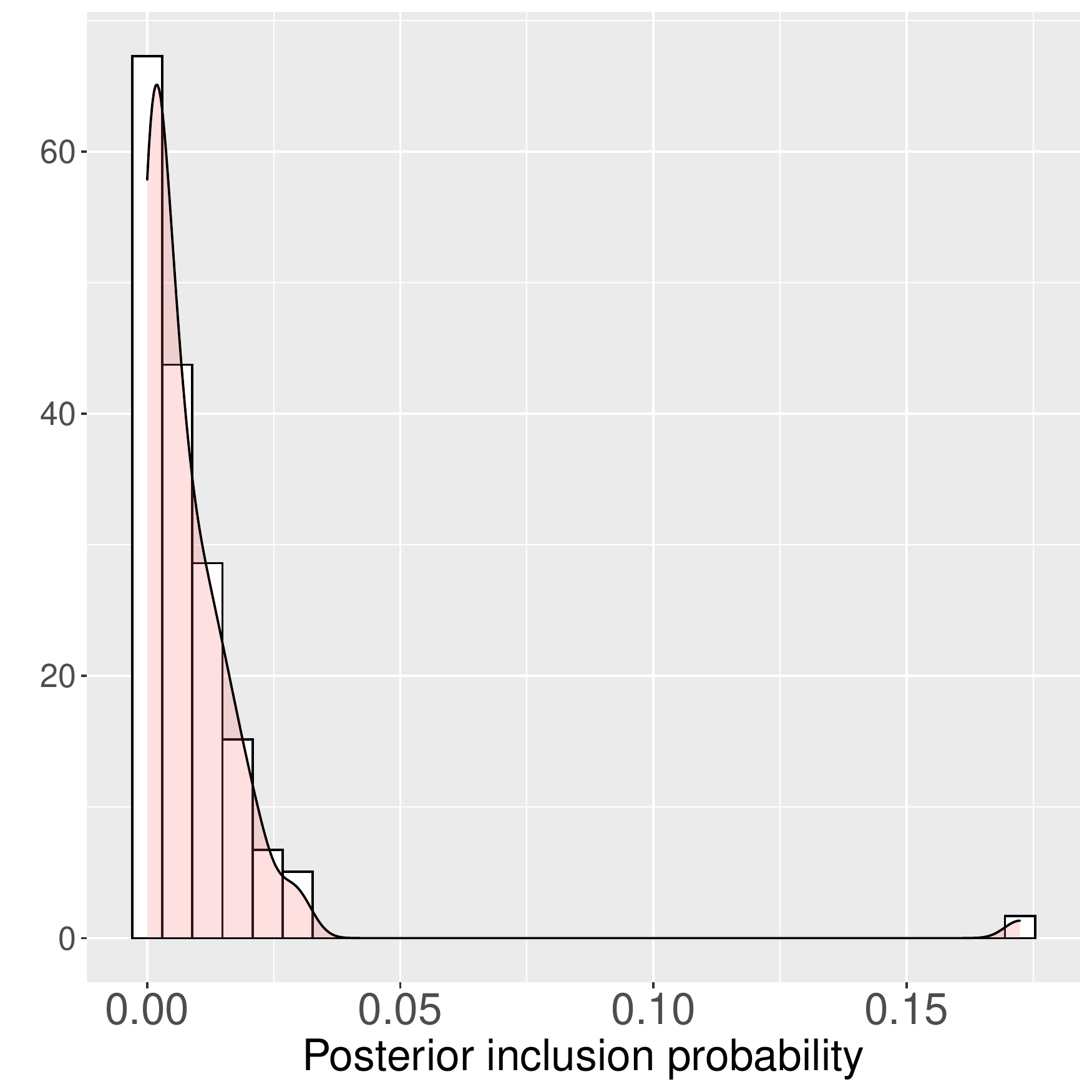} &
\includegraphics[width=.4\textwidth]{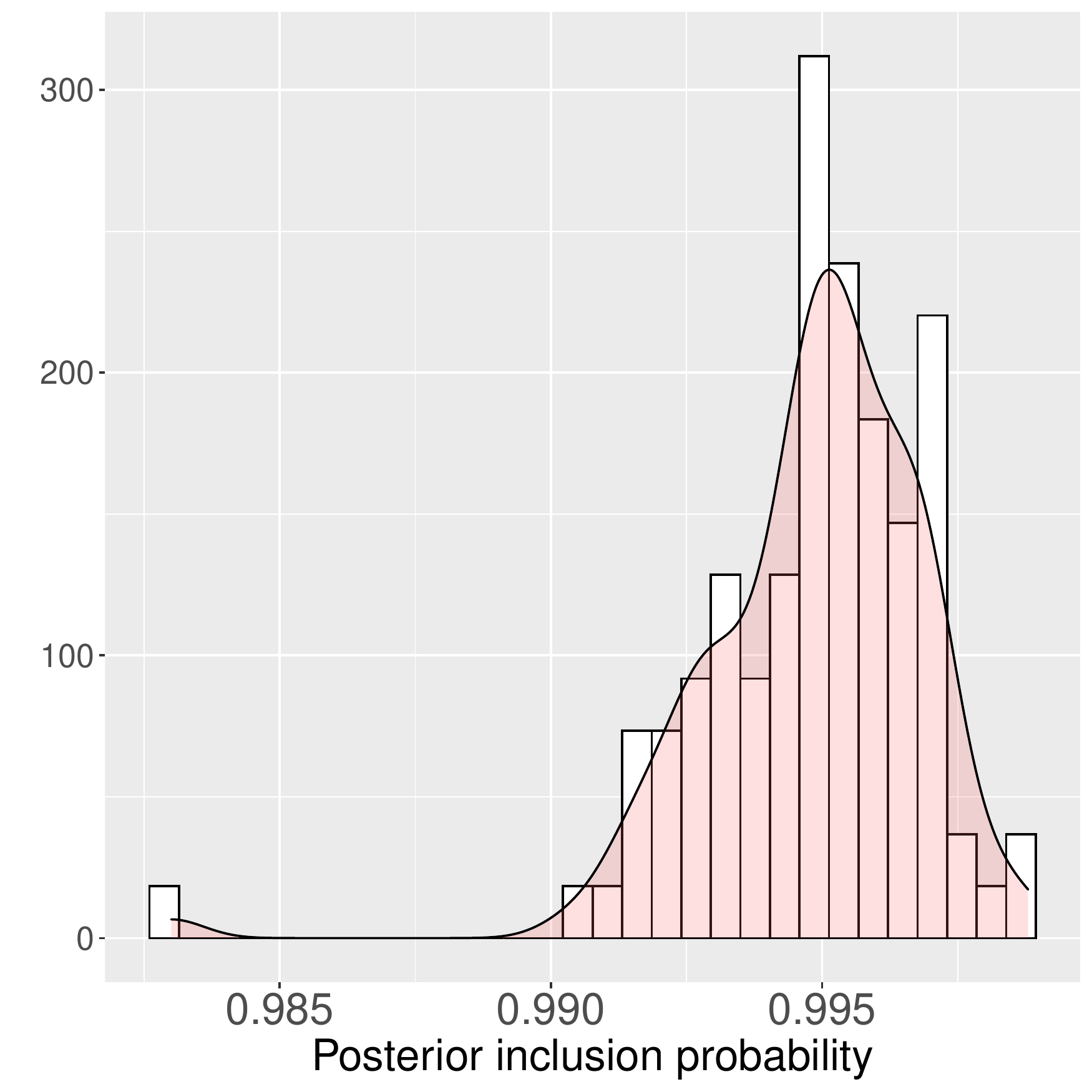} 
\end{tabular}
\caption{Histogram of the estimated posterior probability of the event $\delta = 1$,
called ``posterior inclusion probability."
(a) Data are generated from the Rasch model with $\gamma = 0$.
%the model selection approach selected in at least 95\% of 5,000 iterations the Rasch model.
(b) Data are generated from the latent space model with $\gamma = 1.7$.
%the model selection approach selected in at least 99\% of 5,000 iterations the latent space model.
%the model selection approach selects the data-generating model with high posterior probability.
% under the Rasch model (a) and under the proposed model (b). When the Rasch model was the true model (a), the probability of selecting the proposed model was less than 5\% over 5000 iterations in the 99\% of the simulated datasets. When the proposed model was the true model (b), the probability of  selecting the proposed model was greater than 99\% over 5000 iterations in the 99\% of the simulated datasets. 
%\textcolor{red}{N=200, I=14}
}\label{fg:spike}
\end{figure}

Figure \ref{fg:spike} shows a histogram of the estimated posterior probability of the event $\delta = 1$.
In Figure \ref{fg:spike}(a),
data are generated from the Rasch model with $\gamma = 0$. 
The posterior probability of $\delta=1$ is smaller than 0.05  in at least 99\% of the simulated data sets. 
In Figure \ref{fg:spike}(b),
data are generated from the latent space model with $\gamma = 1.7$.
Here,
the posterior inclusion probability of $\delta=1$ is greater than 0.99 in at least 99\% of the simulated data sets. 
%model selection approach selects the data-generating latent space model in at least 99\% of the simulated data sets. 
If the Rasch model is chosen when the estimated posterior probability of $\delta=1$ is less than $0.5$ and otherwise the latent space model is chosen,
then the data-generating model is selected in all simulated data sets.

These simulation results provide reasonable evidence that the proposed model selection approach helps determine whether the Rasch model with $\gamma=0$ or the latent space model with $\gamma>0$ generated a given dataset.
In other words,
the model selection approach helps decide whether the Rasch model suffices or whether there are systematic deviations from the Rasch model due to unobserved respondent-item interactions,
which the latent space model can capture and represent in a low-dimensional space.
}

\section{Discussion}

{\color{black}
\subsection{Summary}

We have introduced a novel approach to modeling item response data,
capturing deviations from the Rasch model in the form of respondent-item interactions.
While the Rasch model is a classic model and may be a natural starting point in practice,
many item response datasets can be expected to exhibit deviations from the respondent and item effects of the Rasch model,
which implies that there are unexplained interactions between items and respondents.
We have presented evidence of interactions in two empirical examples, 
but in a number of other datasets we tested, 
we likewise observed that interactions among respondents and items are present and non-negligible.
We propose to capture deviations from the Rasch model in the form of respondent-item interactions by embedding both respondents and items in a low-dimensional latent space,
which represents interactions between items, between respondents, and between items and respondents that are not explained by the Rasch model.
The proposed latent space approach has technical advantages over conventional IRT modeling approaches,
because it makes weaker independence and conditional assumptions than conventional approaches,
such as the Rasch model.
An additional, 
intriguing advantage is that it produces a geometrical representation of items and respondents that can provide important insights into how respondents perform on test items.

\subsection{Some final thoughts on practical advantages and possible applications}

We mention here some final thoughts on practical advantages and possible applications of the proposed latent space approach.  
First and foremost,
if the model selection approach described in Section \ref{sec:model_selection} determines that $\gamma > 0$,
then the data exhibit systematic deviations from the main effects of the Rasch model,
that is, 
respondent-item interactions.
The estimated latent space supplies an interaction map that represents those deviations in a low-dimensional space,
providing diagnostic feedback on items as well as respondents. 
For example,
the estimated latent space may be useful for assessing whether test items are differentiated or grouped together as blueprinted by test developers:
e.g., 
the DRV test was developed based on three design factors,
and we found that one design factor (the Presentation of Antecedent) barely contributed to item differentiation and could be dropped without much loss.

In addition,
the estimated latent space could help detect unintended or undesirable forms of test-taking behavior. 
For instance, 
suppose that a computer-based cognitive test with a time constraint is administered (without permission to skip items) and the estimated latent space reveals that a group of respondents is located close to the last test items in the latent space.
That may be an unintended consequence of the fact that most test takers ran out of time and did not respond to the last test items,
so that the few respondents who did respond to them are close to those items in the latent space.
It goes without saying that such conclusions need to be accompanied by additional evidence (e.g., item response times),
but the latent space approach can nonetheless be helpful for diagnosing problems in the first place.

Last,
but not least,
the proposed latent space approach is useful for providing feedback on the test performance of individual test takers or subgroups of test takers.
For example,
in the DRV example, 
we have demonstrated that one can identify items that individual test takers may be struggling with.
Such information could guide classroom instruction,
and help evaluate and improve intervention programs. 

}

\bibliographystyle{apacite}

\bibliography{reference}

\end{document}